\documentclass[onecolumn,nofootinbib,nobibnotes,notitlepage]{revtex4-2}

\newcommand{\um}{\text{\textmu m}}		
\newcommand{\ie}{i.\,e.}				
\newcommand{\eg}{e.\,g.}				

\usepackage{hyperref}
\usepackage{enumerate}
\usepackage{amsmath}
\usepackage{cleveref}
\usepackage{xcolor}
\usepackage{multirow}   
\usepackage{booktabs}	
\usepackage{footnote}

\usepackage[english]{babel} 
\usepackage[utf8]{inputenc}
\usepackage{mathrsfs}
\usepackage{amsmath,amssymb,amsfonts}
\usepackage{graphicx}
\usepackage{array,multirow}
\usepackage{adjustbox}
\usepackage{dsfont}
\usepackage{textcomp}

\crefname{equation}{Eq.}{Eqs.}
\crefname{figure}{Fig.}{Figs.}
\crefname{section}{Sec.}{Secs.}
\crefname{table}{Tab.}{Tabs.}
\crefname{appendix}{Appx.}{Appx.}
\Crefname{equation}{Equation}{Equations}
\Crefname{figure}{Figure}{Figures}
\Crefname{section}{Section}{Sections}
\Crefname{table}{Table}{Tables}
\Crefname{appendix}{Appendix}{Appendices}

\begin{document}
\title{Scattered Light Imaging: Resolving the substructure of nerve fiber crossings\\ in whole brain sections with micrometer resolution}

\author{Miriam Menzel$^1$}
\author{Jan André Reuter$^1$}
\author{David Gr\"a{\ss}el$^1$}
\author{Mike Huwer$^1$}
\author{Philipp Schl\"omer$^1$}
\author{Katrin Amunts$^{1,2}$}
\author{Markus Axer$^1$}

\affiliation{$^{1}$Institute of Neuroscience and Medicine (INM-1), Forschungszentrum Jülich, Wilhelm-Johnen-Straße, 52425 Jülich, Germany\\
	$^{2}$Cécile and Oskar Vogt Institute for Brain Research, University Hospital Düsseldorf, University of Düsseldorf, Moorenstraße 5, 40225 Düsseldorf, Germany}

\email{m.menzel@fz-juelich.de}


\begin{abstract}
	
For developing a detailed network model of the brain based on image reconstructions, it is necessary to spatially resolve crossing nerve fibers. The accuracy hereby depends on many factors, including the spatial resolution of the imaging technique. 3D Polarized Light Imaging (3D-PLI) allows the three-dimensional reconstruction of nerve fiber tracts in whole brain sections with micrometer in-plane resolution, but leaves uncertainties in pixels containing crossing fibers. 
Here we introduce \textit{Scattered Light Imaging (SLI)} to resolve the substructure of nerve fiber crossings. The measurement is performed on the same unstained histological brain sections as in 3D-PLI. By illuminating the brain sections from different angles and measuring the transmitted (scattered) light under normal incidence, light intensity profiles are obtained that are characteristic for the underlying brain tissue structure. We have developed a fully automated evaluation of the intensity profiles, allowing the user to extract various characteristics, like the individual directions of in-plane crossing nerve fibers, for each image pixel at once. 
We validate the reconstructed nerve fiber directions against results from previous simulation studies, scatterometry measurements, and fiber directions obtained from 3D-PLI. 
We demonstrate in different brain samples (human optic tracts, vervet monkey brain, rat brain) that the 2D fiber directions can be reliably reconstructed for up to three crossing nerve fiber bundles in each image pixel with an in-plane resolution of up to 6.5\,\um. 
We show that SLI also yields reliable fiber directions in brain regions with low 3D-PLI signals coming from regions with a low density of myelinated nerve fibers or out-of-plane fibers. This makes Scattered Light Imaging a promising new imaging technique, providing crucial information about the organization of crossing nerve fibers in the brain.

\end{abstract}

\maketitle


\section{Introduction}

The human brain consists of about 100 billion neurons \cite{herculano2009}. To this day, neuroscientists are working on disentangling this gigantic, densely grown network of nerve fibers. 
So far, \textit{diffusion magnetic resonance imaging (dMRI)} is the only possibility to map nerve fiber pathways in-vivo. In postmortem brains, dMRI achieves resolutions down to a few hundred micrometers \cite{roebroeck2018,calabrese2018}. At the same time, many nerve fibers in the white matter have diameters in the order of 1\,\um\ \cite{liewald2014,aboitiz2003}, which creates significant challenges in reconstructing fiber pathways at the level of single nerve fibers, and solving the problem of crossing nerve fibers within the given spatial resolution.
Due to an insufficient knowledge about nerve fiber crossings, fiber tractography algorithms may yield false-positive nerve fiber pathways \cite{maierhein2017}. The microscopy technique \textit{3D-Polarized Light Imaging (3D-PLI)} represents a powerful optical approach to determine the three-dimensional course of nerve fibers in whole, unstained histological brain sections with micrometer in-plane resolution \cite{MAxer2011_1,MAxer2011_2}. The technique has been applied, for example, to reveal previously unknown insights into the connectional architecture of the hippocampal region \cite{zeineh2016}, and it served as cross-validation for fiber tractography algorithms and to improve the interpretation of clinical dMRI data \cite{caspers2019,zeineh2016}. The in-plane resolution reaches a pixel size down to 1.33\,\um\ and allows the mapping of single nerve fiber tracts. Although the course of crossing nerve fibers can be visually tracked in many regions due to a high contrast of the 3D-PLI images, individual nerve fiber orientations cannot be automatically extracted. 3D-PLI yields a single fiber orientation for each image pixel, which creates uncertainties in fiber orientation if the 60\,\um\ thick brain section at this point is comprised of several crossing nerve fibers with different orientations. 
One possibility to disambiguate crossing nerve fibers with high resolution (micrometer or even sub-micrometer range) might be to study the anisotropy of X-ray diffraction patterns with 3D scanning small-angle X-ray scattering (proposed by \cite{georgiadis2020}).
However, the resolution is limited by the available scan time and the measurements require special equipment (among others, a synchrotron).

Simulations and scatterometry measurements \cite{menzel2020,menzel2020-BOEx} have shown that light scattering in the optical regime can also be used to determine the directions of crossing nerve fibers: When shining light through a brain tissue sample and measuring the spatial distribution of scattered light behind the sample, the resulting scattering pattern reveals information about the substructure of the sample, such as the individual directions of (crossing) nerve fibers. While X-ray scattering is sensitive to the regular arrangement of myelin layers (with layer thickness $\sim$\,nm), optical scattering is sensitive to regular arrangements of nerve fibers (with diameters $\sim\,$\textmu m). In contrast to X-ray scattering, the scatterometry measurement can be performed with a simple laser. However, raster-scanning of the sample is still required and the resolution is limited by the minimum available beam diameter (there: $\sim 100$\,\textmu m). 
Recently, it has been shown that a reverse setup (shining light under different angles through the sample and measuring the transmitted light under normal incidence) allows to study the scattering properties of whole brain tissue samples with micrometer in-plane resolution \cite{menzel2020}: Each pixel in the resulting image series contains a light intensity profile that is characteristic for the brain tissue structure at this point and indicates \eg\ the 2D fiber directions of (crossing) nerve fibers. Apart from massively reducing the required measurement time and allowing single-shot images (without rasterizing), this technique has the advantage that the measurement can be performed without dedicated optics (requiring only a standard LED light source and camera).

In this paper, we further develop this technique and introduce \textit{Scattered Light Imaging (SLI)}---an imaging technique that allows the simultaneous extraction of multiple (crossing) nerve fiber directions for whole brain sections with micrometer in-plane resolution. 
While the previous work required the light intensity profiles to be manually evaluated so that only a small number of brain regions could be studied, we now present the open-source software \texttt{SLIX} \textit{(Scattered Light Imaging ToolboX)} that allows a fully automated evaluation of the SLI measurement and the generation of human-readable parameter maps containing various tissue information.
A major improvement is the development of a correction procedure that increases the accuracy of the determined nerve fiber directions from $\pm15^{\circ}$/2 to $\pm2.4^{\circ}$.
In addition, we provide for the first time a validation of the technique by comparing the measured SLI profiles to results obtained from simulations, scatterometry, and 3D-PLI. 
While the previous study considered only two crossing sections of optic tracts and selected regions in a coronal vervet brain section, we present a thorough study on tissue phantoms with up to three crossing sections of optic tracts, a rat brain section, and both coronal and sagittal sections of a vervet monkey brain, including a study on out-of-plane nerve fibers and a long-term study investigating the effect of different embedding times.
We show that SLI yields reliable 2D fiber directions not only in regions with in-plane crossing nerve fibers, but also in other regions with low 3D-PLI signals, like regions with low densities of myelinated nerve fibers or some regions with out-of-plane fiber bundles.


\section{Materials and methods}


\subsection{Preparation of brain sections}
\label{sec:preparation}
The measurements were performed on brain sections obtained from a Wistar rat (male, 3 months old), two vervet monkeys (male, 1 and 2.4 years old), and a human (female, 74 years old). The brains were removed from the skull within 24 hours after death, fixed in a buffered solution of 4\,\% formaldehyde for several weeks, immersed in 20\,\% glycerin, deeply frozen, and cut with a cryostat microtome (\textit{Polycut CM 3500}, \textit{Leica Microsystems}, Germany).\footnote{All animal procedures have been approved by the institutional animal welfare committee at Forschungszentrum Jülich GmbH, Germany, and are in accordance with European Union guidelines for the use and care of laboratory animals. The vervet monkey brains were obtained when the animals were sacrificed to reduce the size of the colony, where they were maintained in accordance with the guidelines of the Directive 2010/63/eu of the European Parliament and of the Council on the protection of animals used for scientific purpose or the Wake Forest Institutional Animal Care and Use Committee IACUC \#A11-219. Euthanasia procedures conformed to the AVMA Guidelines for the Euthanasia of Animals. The human brain was acquired from the Netherlands Brain Bank, in the Netherlands Institute for Neuroscience, Amsterdam. A written informed consent of the subject is available.}  The rat brain and one vervet brain were sectioned coronally, the other vervet brain was sectioned sagittally, into 60\,\um\ thin sections each. The human optic chiasm (including at least 1\,cm of the optic nerves and 1\,cm of the optic tracts) was removed from the human brain and cut along the fiber tracts of the visual pathway into sections of 60\,\um\ and 30\,\um. To study the behavior of crossing fiber bundles with well-defined crossing angles, the sections of the optic chiasm were split into two parts at the median line and the sections of the optic tracts were manually placed on top of each other under different crossing angles (cf.\ \cref{fig:chiasm-parametermaps}(a)). Each brain sample was mounted onto a glass slide, embedded in a solution of 20\,\% glycerin, cover-slipped, and measured up to one day (3D-PLI) or several days (SLI) after tissue embedding.

The reason why SLI measurements were performed up to several months after tissue embedding, is that we found that the overall scattering of the sample decreases with increasing embedding time (probably due to evaporation of the embedding solution), reducing higher-order scattering and increasing the signal-to-noise ratio for crossing nerve fibers, especially in thick tissue samples (see \ref{sec:long-term}). To increase tissue scattering again for samples with long embedding time, some brain sections were revitalized (freshly embedded in glycerin) after several months. \Cref{tab:table} in \ref{sec:samples} lists the section thickness and the days after tissue embedding/revitalization for each measured brain section. 


\subsection{3D Polarized Light Imaging (3D-PLI)}
The 3D-PLI measurements were performed with a polarizing microscope (\textit{LMP-1}, \textit{Taorad GmbH}, Germany), as described in \cite{MAxer2011_1,MAxer2011_2,menzel2020}: The sample is illuminated by linearly polarized, incoherent light with a wavelength of about 550\,nm. The transmitted light passes a circular analyzer (quarter-wave plate and polarizer) and is recorded by a CCD camera (\textit{Qimaging Retiga 4000R}). During the measurement, the polarizer in front of the sample is rotated by $\rho = \{0^{\circ}, 10^{\circ}, \dots, 170^{\circ}$\}. For each rotation angle $\rho$, the camera records an image of the transmitted light intensity, yielding a series of images with a sinusoidal intensity profile per image pixel. The \textit{transmittance} (transmitted light intensity averaged over all rotation angles) is a measure of the tissue attenuation; strong sources of scattering and absorption, such as myelinated nerve fibers, appear dark. The phase of the signal indicates the in-plane orientation angle of the nerve fibers (\textit{fiber direction} $\varphi$). The amplitude of the signal (\textit{retardation} $\vert\sin\delta\vert$, $\delta \propto \cos^2\alpha$) is a measure of the birefringence strength and decreases with increasing out-of-plane angle of the enclosed nerve fibers (\textit{fiber inclination} $\alpha$, cf.\ \cref{fig:scattering-line-profiles}b) \cite{menzel2015}. 
The pixel size in object space is $1.33 \times 1.33$\,\um$^2$.


\subsection{Scattered Light Imaging (SLI)}
\label{sec:SLI}

\begin{figure}[b]
	\centering
	\includegraphics[width=0.8\textwidth]{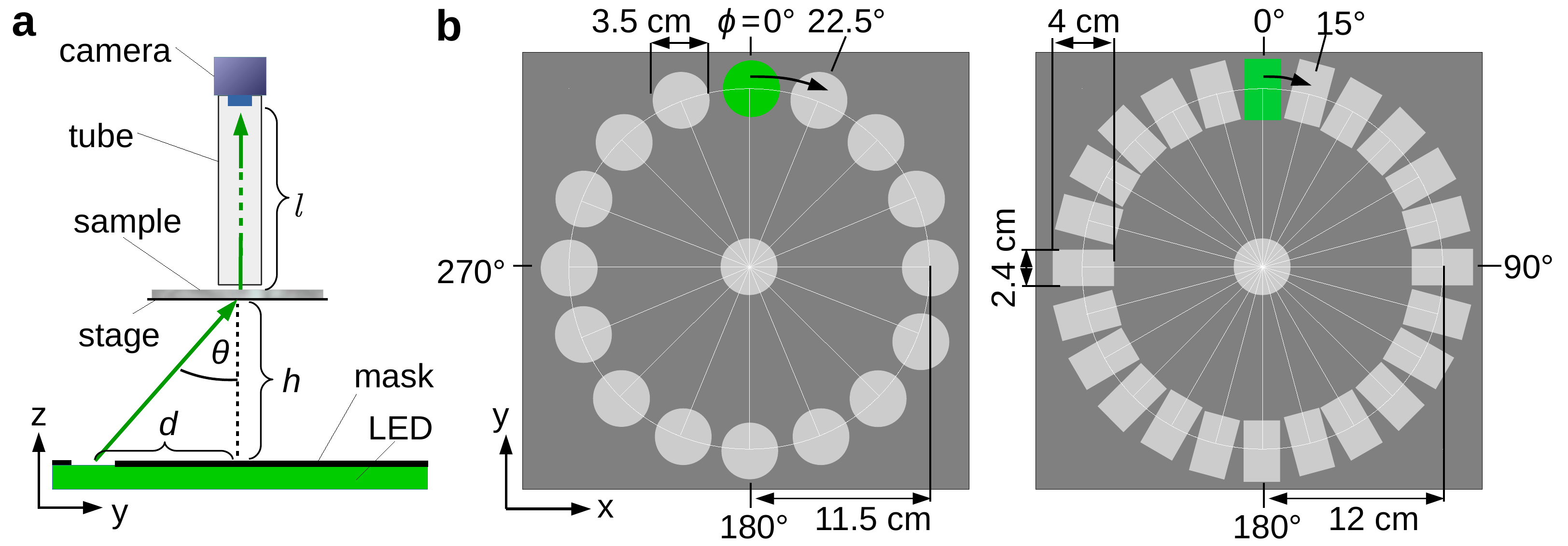}
	\caption{\textbf{Schematic drawing of the setup used for Scattered Light Imaging (SLI).} \textbf{(a)} Side view of the setup consisting of an LED panel (with diffuser plate), a specimen stage with sample, a sheltering tube, and a CCD camera. A mask with a hole is placed on top of the LED panel so that the sample is illuminated under a large polar angle $\theta$. \textbf{(b)} Top view of the two masks used for the SLI measurements (Left: circular holes, Right: rectangular holes).}
	\label{fig:scattering-setup}
\end{figure}

The setup for the SLI measurements (\cref{fig:scattering-setup}) has been introduced by \cite{menzel2020}: The sample is illuminated by mostly unpolarized and incoherent light with a wavelength of about 525\,nm. A mask with a hole is placed on top of the light source (LED panel with diffuser plate) so that the center of the sample is illuminated under a large polar angle $\theta$. The transmitted light intensity under normal incidence is recorded by a CCD camera (\textit{AVT}, \textit{Oscar F-810C}, Germany). To obtain a certain field of view, the distance between light source and sample ($h$) and between sample and camera objective ($l$) can be varied.
The distance between sample and camera objective needs to be large enough to ensure that the scattered light falls mostly vertically onto the camera ($\theta < 3^{\circ}$). Alternatively, an objective lens with small numerical aperture or a telecentric lens with small iris size can be used.
To avoid the detection of ambient light, the light path between sample and camera objective is sheltered by a dark tube.
During the SLI measurement, the mask with the hole is rotated, starting at the twelve o'clock position ($\phi = 0^{\circ}$) and rotating clock-wise. For each illumination angle (azimuthal angle $\phi$), an image is recorded by the camera. Each pixel in the resulting image series contains a light intensity profile (\textit{SLI profile}, $I(\phi)$), which is characteristic for the measured brain region and can be used to compute \eg\ the direction angles of the nerve fibers. To enable a better comparison between measurements, the intensity values for each illumination angle $\phi$ are divided by the average intensity over all illumination angles for each image pixel, yielding a \textit{normalized} SLI profile $I_{\rm N}(\phi)$. (The evaluation of the SLI profiles is described in \cref{sec:evaluation-line-profiles} in more detail.)

The measurements presented here were performed with two different types of masks (see \cref{fig:scattering-setup}(b)): one mask with circular holes with distance $d=11.5$\,cm to the center, diameters of 3.5\,cm, and steps of 22.5$^{\circ}$ (left)---and another mask with rectangular holes with distance $d=12$\,cm to the center, sizes of $2.4 \times 4.0$\,cm$^2$, and steps of 15$^{\circ}$ (right). 
If not otherwise stated, the measurements were performed with the mask with rectangular holes and $15^{\circ}$-steps.
Some regions of interest were measured in $5^{\circ}$-steps using rectangular holes with sizes of $0.4 \times 4.0$\,cm$^2$.
The circular hole in the middle of the masks was used as a reference in order to align mask, probe, and camera, and to adjust the field of view of the camera in the bright field. The region of interest was illuminated by the middle hole to guarantee a constant polar angle of illumination during the measurement:
$\theta = \arctan(d/h) \approx 46$--$48^{\circ}$. Depending on the desired field of view, an objective lens with 50\,mm or 90\,mm focal length (\textit{Rodenstock}, \textit{Apo-Rodagon-N50}/\textit{90}) was used, with a focal ratio of 4. The distance between sample and camera objective was varied ($l=19.3$--$34.8$\,cm), yielding pixel sizes in object space between 6.5\,\um\ and 15.0\,\um.
To reduce noise, four images were taken for each illumination angle $\phi$ and averaged.
\Cref{tab:table} lists the settings of the SLI measurements (mask, sample/camera distances, objective lens, exposure time, pixel size) for all measured samples.
To uniquely identify each measurement (also for measurements of the same sample at different times), a unique identifier (\eg\ \#123) was assigned to each measurement.


\subsection{Image registration}
\label{sec:registration}

To compare different measurements of the same tissue sample, the average transmitted light intensities were registered onto each other using in-house software tools based on the software packages \cite{ITK}, ELASTIX \cite{elastix}, and ANTs \cite{avants2008,avants2011,shamonin2013}, which perform linear and non-linear transformations. The same transformation was applied to other corresponding images and modalities. 


\subsection{Evaluation of SLI profiles}
\label{sec:evaluation-line-profiles}

\begin{figure}[h!]
	\centering
	\includegraphics[width=\textwidth]{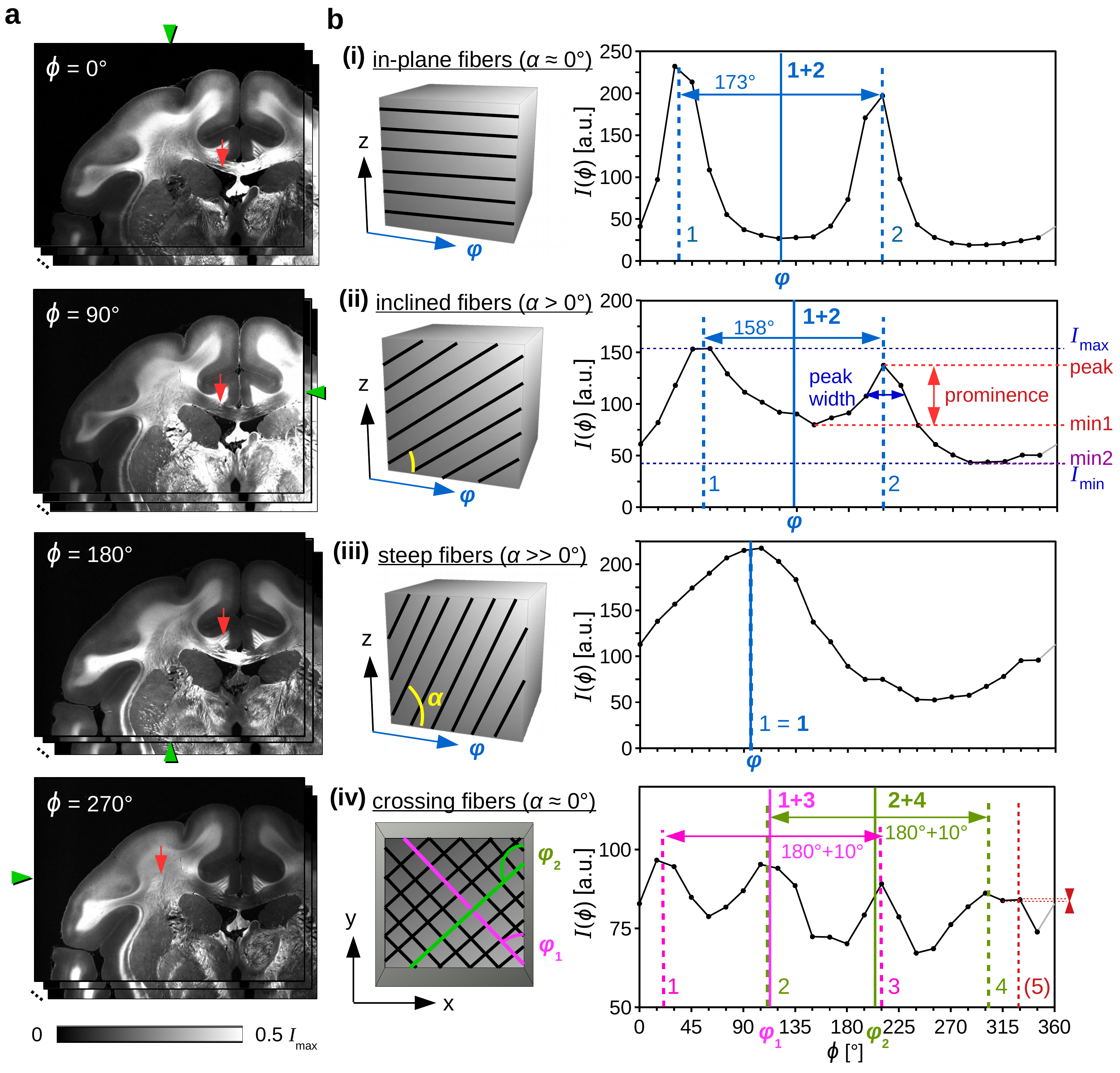}
	\caption{\textbf{Evaluation of SLI profiles.} \textbf{(a)} Image series obtained from an SLI measurement of a coronal vervet brain section (upper left corner, see \cref{tab:table}\#90). The measurement was performed using the mask with rectangular holes and $15^{\circ}$-steps. The green arrowheads at the image borders indicate the illumination angle $\phi$. \textbf{(b)} Left: Schematic drawing of the nerve fiber geometries associated with the evaluated brain regions (in-plane, inclined, steep, and in-plane crossing fibers). The angle $\varphi$ denotes the fiber direction in the xy-plane, the angle $\alpha$ the out-of-plane inclination. Right: Corresponding SLI profiles obtained from a region of $10 \times 10$ pixels, indicated by the red arrows in (a). The prominence of the peaks (in red) is computed by the vertical distance between the top of the peak and the higher of the two neighboring minima, min1 and min2 (minimum values between the peak and the next higher points on the left and right side of the peak). Only peaks with a prominence $\geq$ 8\,\% are considered for evaluation (the fifth peak in (iv) is assumed to be a noise artifact). The peak width (dark blue) is determined as the full width of the peak at a height corresponding to the peak height minus half of the peak prominence. The dashed vertical lines indicate the corrected positions of the determined peaks, the solid vertical lines the computed fiber direction angles $\varphi$.}
	\label{fig:scattering-line-profiles}
\end{figure}

Preliminary measurements by \cite{menzel2020} have shown that the SLI profiles are characteristic for different nerve fiber constellations (see \cref{fig:scattering-line-profiles}): Parallel in-plane nerve fibers yield an SLI profile with two distinct peaks that lie approximately $180^{\circ}$ apart (i). With increasing fiber inclination $\alpha$, the two peaks move closer together until they merge in one broad peak (iii). The middle position between the two peaks indicates the fiber direction $\varphi$, \ie\ the 2D in-plane projection of the actual 3D fiber orientation.\footnote{The term ``fiber direction'' will be used in the following to refer to the xy-component (2D in-plane projection) of the 3D nerve fiber orientation.} If two in-plane fiber bundles cross each other, the resulting SLI profile shows four distinct peaks. Each pair of peaks (lying approximately $180^{\circ}$ apart) belongs to one fiber bundle, respectively (iv).
Hence, by investigating the peaks of the measured SLI profiles, it is possible to obtain information about the out-of-plane angles and---more importantly---the in-plane crossing angles of the nerve fibers in an investigated brain section. 

The positions of the peaks (local maxima) were computed with \textit{Python} using the \textit{SciPy} package (version 1.5.2) and the function \texttt{scipy.signal.find\_peaks} \cite{scipy}, taking the $360^{\circ}$-periodicity of the signal into account.
To avoid that small irregularities in the SLI profiles are detected as peaks, only significant peaks with a certain \textit{prominence}  (\texttt{scipy.signal.peak\_prominences}) were considered as peaks. The prominence is the vertical distance between the top of the peak and the higher of the two neighboring minima (see \cref{fig:scattering-line-profiles}(b)(ii) in red).
If not otherwise stated, only \textit{prominent} peaks, \ie\ peaks with a prominence $\geq 8$\,\% of the total signal amplitude ($I_{\rm max} - I_{\rm min}$) were considered, see \ref{sec:prominence-determination} for derivation.
To account for inaccuracies introduced by the discretization of the SLI profiles, the determined peak positions were corrected by calculating the geometric center of the peak tips (with a height corresponding to 6\,\% of the total signal amplitude, see \ref{sec:peak-positions}).
With this correction method, the peak positions in SLI profiles with 15$^{\circ}$-steps can be determined with a Gaussian discretization error of $\pm 2.4^{\circ}$ (see \cref{fig:PeakPos-Correction}), which is much less than without correction (approx.\ uniform distribution between $\pm 15^{\circ}/2$).

The fiber direction angles $\varphi$ (see solid vertical lines in blue/magenta/green in \cref{fig:scattering-line-profiles}(b)) were computed from the corrected peak positions (dashed vertical lines in corresponding colors): 
In regions with one prominent peak (iii), the fiber direction was defined as the peak position itself (1). In regions with two prominent peaks ((i) and (ii)), the fiber direction was computed by the average position between the two peaks ($1+2$). In regions with four or six prominent peaks with pair-wise distances of $180^{\circ} \pm 35^{\circ}$, the direction angles were computed by the mid positions of the peak pairs (as shown in (iv)). In all other cases, the fiber directions were not computed.

To enable a fully automated evaluation of the SLI profiles for whole brain tissue samples, we developed the open-source software \texttt{SLIX} \textit{(Scattered Light Imaging ToolboX)} which is available on \href{https://github.com/3d-pli/SLIX}{GitHub}. The software evaluates the peaks of all measured SLI profiles and generates different parameter maps---containing information about the number of (prominent) peaks, the average peak prominence, width, and distance, as well as the derived nerve fiber directions \cite{slix}.


\subsection{Comparison to scatterometry profiles}
\label{sec:scatterometry}

For validation, the results from SLI were compared to results from coherent Fourier scatterometry with non-focused, normally incident laser light \cite{menzel2020-BOEx}. 
For the comparison, we used smoothed polar integrals (\textit{scatterometry profiles}) from the data set \cite{dataset}. The scatterometry profiles were obtained by measuring 232 spots in different brain tissue samples (circles in \cref{fig:samples}) with a collimated laser beam with 633 nm wavelength and 1.12\,mm diameter, a numerical aperture of 0.4, and an exposure time of 30\,ms, and recording the Fourier transform of the image plane (scattering pattern).\footnote{Due to the aperture, the scattering pattern contains only angles $\theta \leq \arcsin(0.4) \approx 23^{\circ}$.} The scatterometry profiles represent how strongly the light is scattered in each azimuthal direction (polar integral of the scattering pattern).

From the scatterometry profiles, the position and prominence of the peaks were computed and the determined positions of the peaks (with prominence $\geq 3\,\%$) were compared to the positions of the SLI peaks (with prominence $\geq 8\,\%$) for the same tissue spots (see \ref{sec:samples} for more details). Apart from comparing the number and positions of the peaks, we computed the sum of differences between SLI and scatterometry profiles. For better comparison, the minimum intensity values were subtracted from all profiles and the resulting intensity values were divided by the average signal intensity, respectively, yielding normalized line profiles $I_{\rm N}(\phi)$.


\subsection{Comparison to simulated scattering patterns}
\label{sec:simulation}

In regions with out-of-plane nerve fibers, the SLI profiles were compared to simulated scattering patterns by \cite{menzel2020}. 
The simulations were performed by modeling the propagation of a plane wave with coherent, circularly polarized light ($\lambda = 550$\,nm) and normal incidence through artificial nerve fiber constellations, using a finite-difference time-domain algorithm. The nerve fibers were modeled by an inner axon and a surrounding myelin sheath with different refractive indices and uniformly distributed fiber diameters between 1.0--1.6\,\um, in a volume of $30 \times 30 \times 30$\,\um$^3$ (cf.\ \cref{fig:vervet-inclined-fibers}(c), left). The scattering patterns (cf.\ \cref{fig:vervet-inclined-fibers}(c), right) were obtained by computing the intensity distribution of the scattered light wave behind the sample on a hemisphere projected onto the xy-plane.
To take the finite size of the illumination beam in SLI into account, the simulated scattering patterns were blurred with a Gaussian filter with a diameter corresponding to the diameter of the circular hole in the mask ($8^{\circ}$ in angular space), and sampled along a circle corresponding to $\theta = 48^{\circ}$ (polar angle of illumination used in SLI), yielding a simulated line profile.


\section{Results}
\enlargethispage{0.5cm}


\subsection{Comparison of SLI and scatterometry measurements}
\label{sec:SLI-scatterometry-comparison}

To validate the results obtained from Scattered Light Imaging, the (normalized) SLI profiles were compared to scatterometry profiles of the same tissue spots. For comparison, 200 tissue spots in two/three crossing sections of human optic tracts and in coronal/sagittal vervet brain sections were selected (see non-white circles in \cref{fig:samples}(a)). Tissue spots that do not lie completely inside the tissue or that belong to anatomically known brain regions with out-of-plane nerve fibers were not used for comparison.\footnote{The scattering patterns of out-of-plane fibers are not expected to be radially symmetric (cf.\ \cite{menzel2020}, fig.\ 7(a)) so that the scatterometry profiles (obtained from small scattering angles $\theta \leq 23^{\circ}$) are expected to be very different from the corresponding SLI profiles (with $\theta \approx 48^{\circ}$). The SLI profiles for out-of-plane nerve fibers were therefore only compared to simulated line profiles.}

\begin{figure}[h!]
	\centering
	\includegraphics[width=\textwidth]{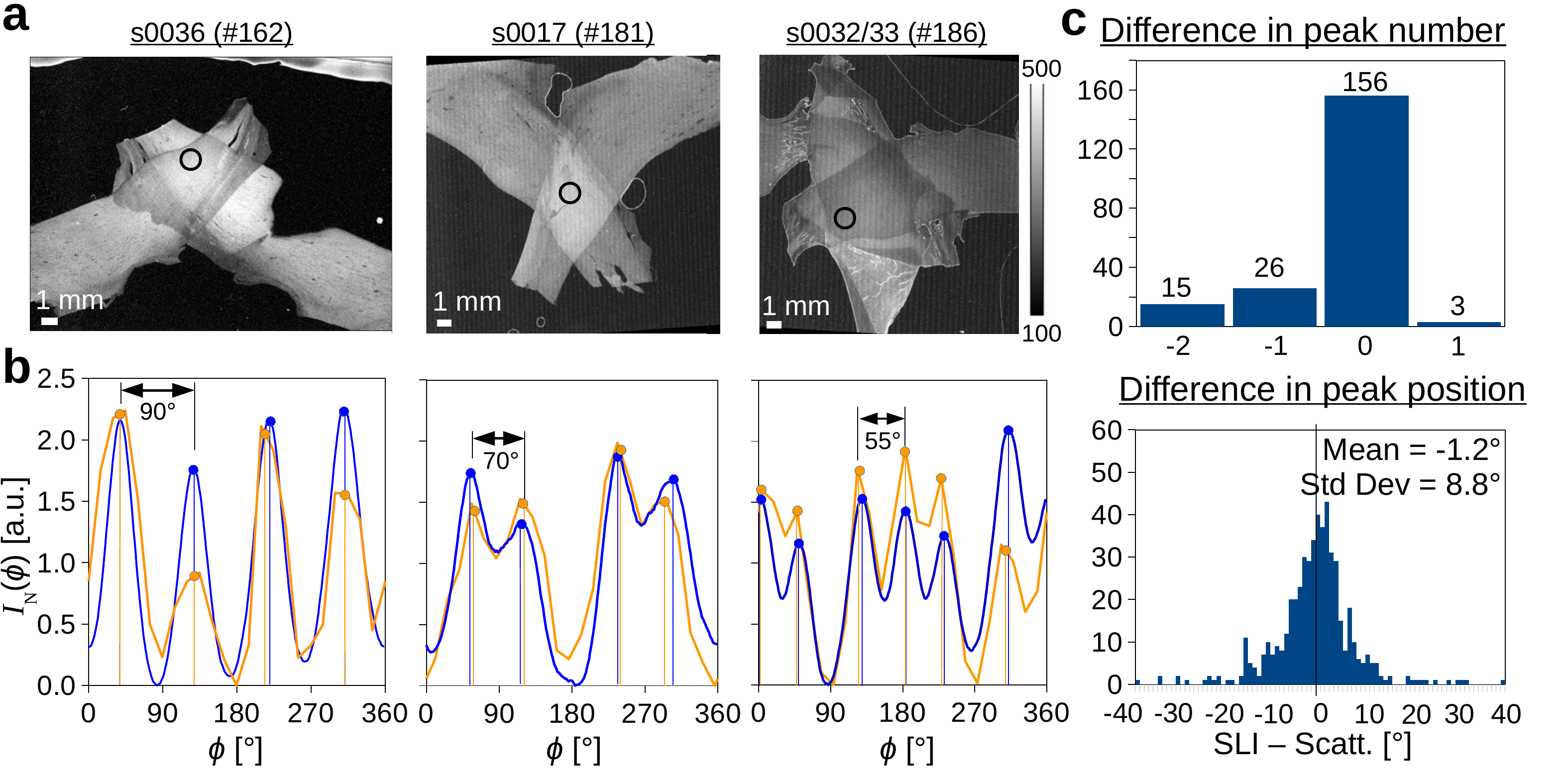}
	\caption{\textbf{Comparison of SLI and scatterometry profiles.} \textbf{(a)} Average transmitted light intensity of the SLI measurement for two and three crossing sections of human optic tracts (see \cref{tab:table}\#162,181,186). \textbf{(b)} Normalized SLI profiles (orange) and scatterometry profiles (blue) obtained from the tissue spots indicated in (a). The vertical lines show the determined peak positions. \textbf{(c)} Difference in the number of detected peaks (top) and in the detected peak positions (bottom) between the SLI and scatterometry profiles for 200 evaluated brain tissue spots (see non-white circles in \cref{fig:samples}).}
	\label{fig:SLI_vs_scatterometry_comparison}
\end{figure}

\Cref{fig:SLI_vs_scatterometry_comparison}(b) shows the normalized SLI profiles (orange) and the scatterometry profiles (blue) for two and three crossing sections of optic tracts with crossing angles $90^{\circ}$, $70^{\circ}$, and $55^{\circ}$ (the evaluated tissue spots are marked in (a)). Despite some differences in shape, the line profiles correspond very well to each other and the positions of the peaks (indicated by vertical lines) are almost identical.

The histograms in \cref{fig:SLI_vs_scatterometry_comparison}(c) show the difference in peak number and position for all spots.
The vast majority of the line profiles (78\,\%) show the same number of peaks for SLI and scatterometry. For 22\,\% of the SLI profiles, one or two peaks less are detected (due to the $15^{\circ}$-discretization of the SLI profiles). The difference between the SLI and scatterometry peak positions is: $-1.2^{\circ}\pm 8.8^{\circ}$.

\Cref{fig:samples}(c) shows the sum of distances between the SLI and scatterometry profiles. In very inhomogeneous regions with crossing fibers, like the corona radiata of the vervet brain (magenta circles in \cref{fig:samples}(a)), the sum of distances is especially large. Possibly, the tissue spots could not be located at exactly the same position or details in the fiber structure changed between measurements.

Taking into account that the SLI measurement is completely different to the scatterometry measurement, and that it has been performed several weeks beforehand, the results show that the SLI profiles are compatible with those obtained from scatterometry, which validates the SLI approach.


\subsection{Optic tracts as model system for in-plane crossing nerve fiber bundles}
\label{sec:chiasm}

\begin{figure}[htbp]
	\centering
	\includegraphics[width=0.7\textwidth]{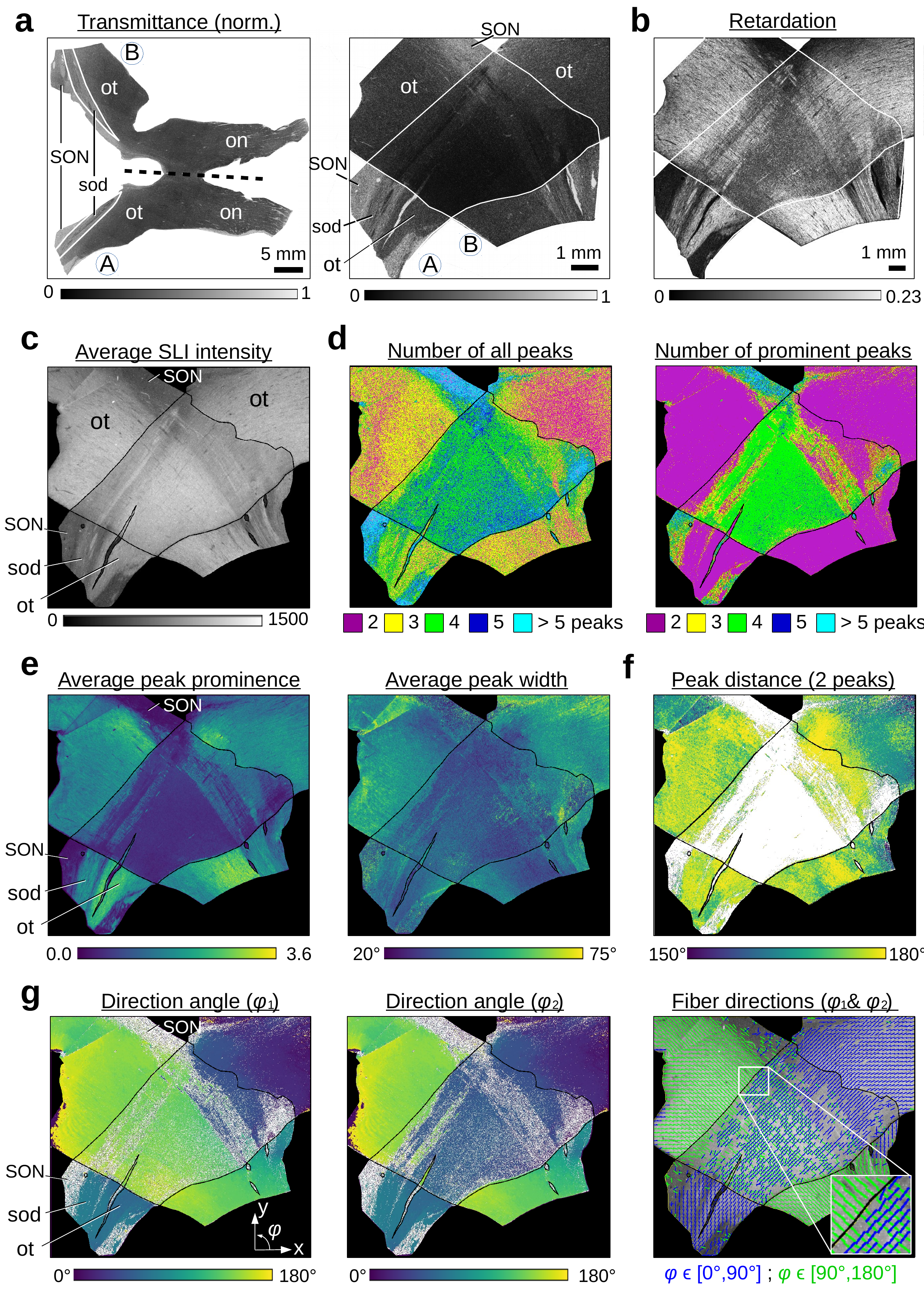}
	\caption{\textbf{Parameter maps from a 3D-PLI and an SLI measurement of two crossing sections of human optic tracts} (section 36, \cref{tab:table}\#158). For better distinction, the crossing sections are delineated by contour lines. 
		\textbf{(a)} Normalized transmittance images for the whole section of the human optic chiasm (on the left) and for the crossing sections of optic tracts (on the right). The chiasm section was divided along the dashed line into two parts, and the sections of optic tracts were placed on top of each other (B on top of A). (on = optic nerve, ot = optic tract, SON = supraoptic nucleus, sod = supraoptic decussation). \textbf{(b)} Retardation image of the crossing sections of optic tracts. \textbf{(c)} Average intensity of the (non-normalized) SLI profiles. \textbf{(d)} Number of peaks in the SLI profiles (left:\ all peaks with prominence $\geq 0\,\%$; right:\ only peaks with prominence $\geq 8\,\%$ of the total signal amplitude), shown in different colors. \textbf{(e)} Average peak prominence and average peak width (computed by averaging over all prominent peaks in an SLI profile). The minimum values are displayed in dark blue, the maximum values in yellow. \textbf{(f)} Distance between two prominent peaks in the SLI profiles (regions with more than two prominent peaks are shown in white). \textbf{(g)} Fiber direction angles ($\varphi_1$, $\varphi_2$) computed from the arithmetic mean value of the peak pair positions for regions with two or four prominent peaks in the SLI profiles. Regions with two prominent peaks (interpreted as non-crossing fibers) are shown in both images, $\varphi_1$ and $\varphi_2$. The image on the right shows the 2D fiber directions as blue and green lines sampled from every 24$^{\rm th}$ image pixel.}
	\label{fig:chiasm-parametermaps}
\end{figure}

To test the accuracy of SLI on a model system with well-known fiber orientations, we used two and three crossing sections of optic tracts (generated by dividing sections of a human optic chiasm along the median line and placing the sections of optic tracts manually on top of each other under different crossing angles, cf.\ \cref{fig:chiasm-parametermaps}(a)). The optic tracts were used as model systems because they contain many parallel (myelinated) nerve fibers with well-defined orientations and have already shown their value in previous work \cite{menzel2020,menzel2020-BOEx}.
The model system can be used to study up to three in-plane crossing fiber bundles with different crossing angles.\footnote{Producing a model system with more than three crossing fiber bundles would be very challenging, because the more sections are used, the thinner the sections must be (to maintain transparency) and the more difficult it becomes to obtain a sample with defined crossing angles (the sections move when being placed on top of each other). Also, it is only possible to study in-plane nerve fibers because cutting the optic tract under a certain angle would yield samples with unknown fiber inclinations after being placed in between the glass slides, and inclining the sample itself by using a tiltable specimen stage would not yield comparable results to inclined nerve fibers in real brain sections, where the fibers are cut diagonally instead of running parallel to the glass plates.}


When studying the SLI profiles of regions with parallel fibers (single tissue layer) and crossing fibers (double/triple tissue layers), it becomes apparent that the signal in the crossing region is a superposition of the signals in the corresponding single tissue layers (see \ref{sec:fiber-crossings}).
The crossing angles of the fiber tracts can be correctly determined in the crossing region, not only for two crossing sections of optic tracts with $90^{\circ}$ and $70^{\circ}$ crossing angles, but also for three crossing sections with $55^{\circ}$ crossing angle.

The three crossing sections of optic tracts were measured both with $15^{\circ}$- and $5^{\circ}$-steps. 
Although the SLI profiles show more details for $5^{\circ}$-steps, the determined peak positions correspond very well to the peak positions obtained from the SLI profiles with $15^{\circ}$-steps (see \cref{fig:chiasm-3xcross_15-vs-5deg}). To reduce measurement time, all following SLI measurements were performed with $15^{\circ}$-steps.

Two crossing sections of optic tracts were measured both with 3D-PLI\footnote{Note that the 3D-PLI measurement was performed right after tissue embedding, while the SLI measurement was performed several months later, right after revitalization of the tissue, so that the samples show small differences in shape.} and with SLI. \Cref{fig:chiasm-parametermaps} shows the resulting parameter maps. The SLI parameter maps were generated with the \texttt{SLIX} software:
For each image pixel, the corresponding SLI profile was normalized, the peaks were determined, and the peak prominences, peak widths, and peak positions were computed.

The average intensity of the SLI profiles (\cref{fig:chiasm-parametermaps}(c)) is a measure for the overall scattering of the sample. While the \textit{optic tracts (ot)} are highly scattering, the \textit{supraoptic decussation (sod)}, which contains less myelinated nerve fibers, is less scattering. In regions belonging to the \textit{supraoptic nucleus (SON)}, which contains mostly cell bodies, the SLI profiles show a large number of randomly distributed peaks with a small average peak prominence (see \cref{fig:chiasm-parametermaps}(d),(e) on the left) and cannot be reliably evaluated.
When taking all peaks into account (left image in \cref{fig:chiasm-parametermaps}(d)), many regions are not correctly assigned (regions with parallel fibers show three peaks or more; regions with crossing fibers show five peaks ore more).
When considering only prominent peaks (right image in (d)), regions with parallel and crossing fibers are more correctly assigned (parallel fibers with two peaks are shown in magenta, two crossing fiber bundles with four peaks are shown in green), as also found in \ref{sec:prominence-determination}. The average prominence of the peaks is larger in regions with parallel fibers and lower in regions with crossing fibers, the same holds for the average peak width ((e) on the right). The distance between the peaks, evaluated for regions with two prominent peaks (f), lies mostly between $170^{\circ}$ and $180^{\circ}$, showing that most fibers are orientated in-plane, as expected. At the image borders, some values lie between $160^{\circ}$ and $170^{\circ}$, suggesting that the fibers are more inclined there.
The direction maps in (g) show the two fiber directions ($\varphi_1$ and $\varphi_2$) computed from the different peak pairs in regions with two and four prominent peaks. The individual nerve fiber directions in the two sections of the optic tracts are clearly visible in the crossing region, and transition smoothly to the regions with parallel fibers (single tissue layer). The image on the right shows the 2D fiber directions as blue and green lines computed from the direction angles (for every 24$^{\rm th}$ image pixel). The lines nicely show the course of the fibers in the optic tracts, not only for regions with parallel fibers (single tissue layer), but also within the crossing region (double tissue layer).

\begin{figure}[htbp]
	\includegraphics[width=0.9\textwidth]{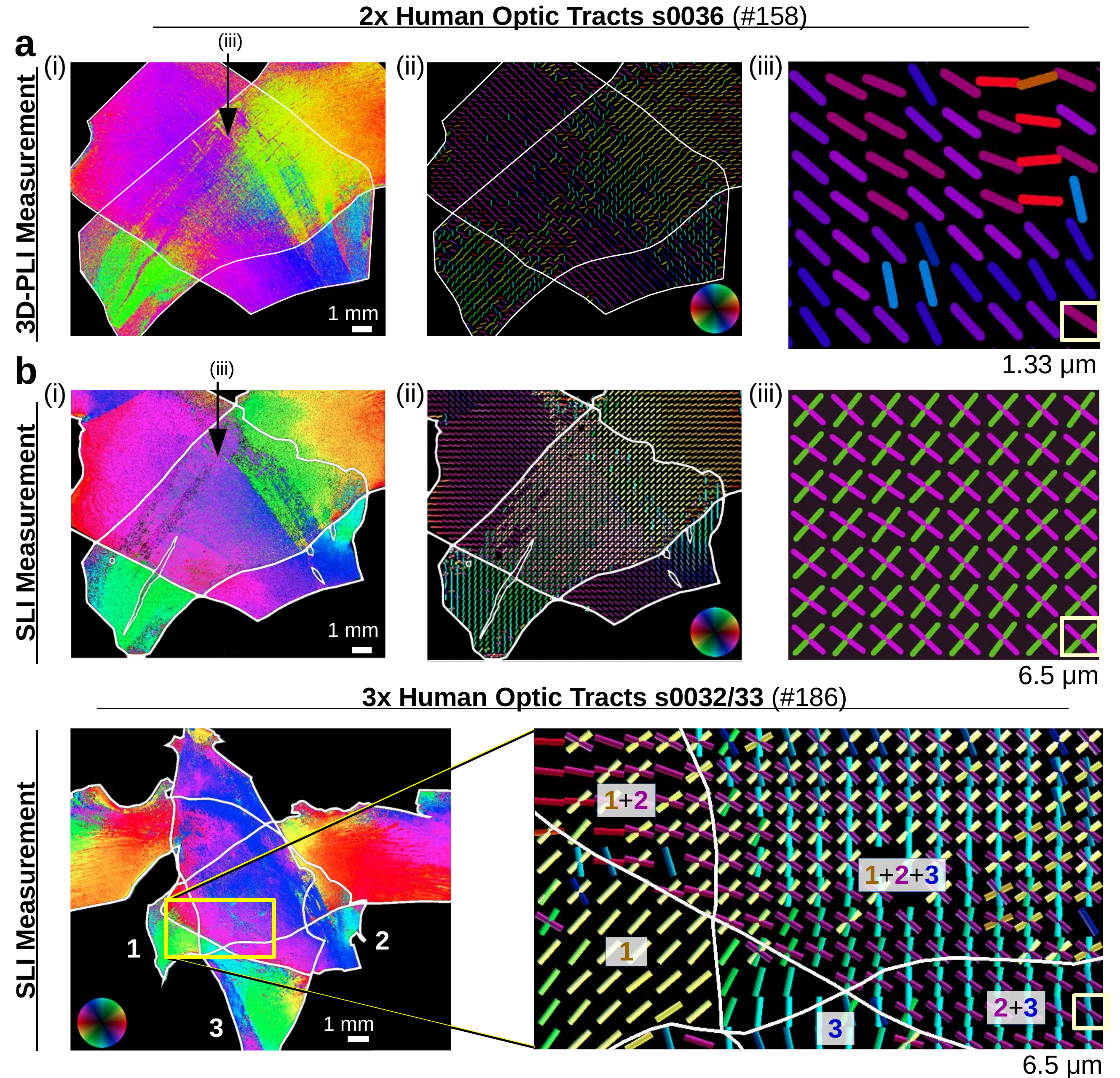}
	\caption{\textbf{Reconstructed nerve fiber directions from 3D-PLI and SLI measurements for two and three crossing sections of human optic tracts.} \textbf{(a--b)} In-plane fiber direction angles obtained from a 3D-PLI measurement and an SLI measurement of two crossing sections of optic tracts (section 36, \cref{tab:table}\#158). 
		The fiber direction is encoded by hue according to the color legend in (ii). The left images (i) show the color-coded fiber directions of each image pixel. For image pixels in which the SLI measurement yields different fiber direction angles, only a single fiber direction angle is displayed. The middle images (ii) show the fiber directions as bars (a) for every 150$^{\rm th}$ image pixel, and (b) for every 30$^{\rm th}$ image pixel, using the same color code. The SLI fiber directions were replaced by the median fiber directions of each $3 \times 3$ pixels beforehand. (In this way, pixels for which the SLI measurement yields no fiber direction are replaced by the main fiber direction of the neighboring pixels. By taking the median and not the average, we ensure that the fiber directions are not distorted by averaging over the different fiber directions in the two bundles.)
		The right images (iii) show a zoom-in of (i) for a small region marked by the arrow, respectively. The corresponding transmittance and retardation images of the 3D-PLI measurement are shown in \cref{fig:chiasm-parametermaps}(a),(b). \textbf{(c)} In-plane fiber direction angles obtained from an SLI measurement of three crossing sections of optic tracts (sections 32/33, \cref{tab:table}\#186). The left image shows the color-coded fiber directions of each image pixel, the right image shows the fiber direction as colored bars for every $40^{\rm th}$ pixel (after computing the median of $3 \times 3$ pixels). For better reference, the sections of optic tracts are delineated by white contour lines and labeled by 1,2,3.
	}
	\label{fig:chiasm_SLI_vs_PLI}
\end{figure}

The fiber directions obtained from SLI were compared to fiber directions obtained from a 3D-PLI measurement of the same tissue sample for different levels of detail (see \cref{fig:chiasm_SLI_vs_PLI}(a--b): (i) whole image with fiber directions displayed for every image pixel, (ii) whole image with fiber directions displayed for every 150$^{\rm th}$ (a) and 30$^{\rm th}$ (b) image pixel, (iii) zoom-in of small region (marked by arrow in (i)) with fiber directions displayed for every image pixel. (High-resolution parameter maps can be found in the data repository \cite{EBRAINS}). 

In contrast to SLI, the 3D-PLI measurement yields only a single nerve fiber orientation for each measured image pixel. In regions with non-crossing nerve fibers (single tissue layer), 3D-PLI and SLI yield similar (2D) nerve fiber directions. In regions with crossing nerve fiber layers, the fiber directions from 3D-PLI correspond to the fiber directions in one or the other fiber layer (magenta/dark blue or yellow/green), and in some regions to an intermediate fiber direction of both (cyan), see \cref{fig:chiasm_SLI_vs_PLI}(a)(ii) and (iii). This observation is consistent with simulations by \cite{dohmen2015} who showed that in regions of two crossing fiber bundles with nearly $90^{\circ}$ crossing angle, the fiber direction obtained from 3D-PLI corresponds to the direction of the fiber bundle that contains $>50\,\%$ of the fibers. For a slightly different crossing angle, the resulting fiber direction is an intermediate direction of the two fiber bundles.

While the 3D-PLI measurement (a) yields only a single nerve fiber orientation, also in regions with crossing fibers, the SLI measurement (b) clearly shows the individual nerve fiber directions of the two crossing optic tracts. 
The SLI measurement reliably reconstructs the individual nerve fiber directions also in regions with three crossing fiber bundles/layers (c). The course of the three individual fiber bundles is clearly visible both in regions with two and three crossing sections of optic tracts (see zoomed-in region on the right).


\subsection{Nerve fiber architectures in whole brain sections}
\label{sec:whole-sections}


\subsubsection{Comparison of SLI and 3D-PLI measurements}
\label{sec:whole_non-crossing}

\begin{figure}[htbp]
	\centering
	\includegraphics[width=0.9\textwidth]{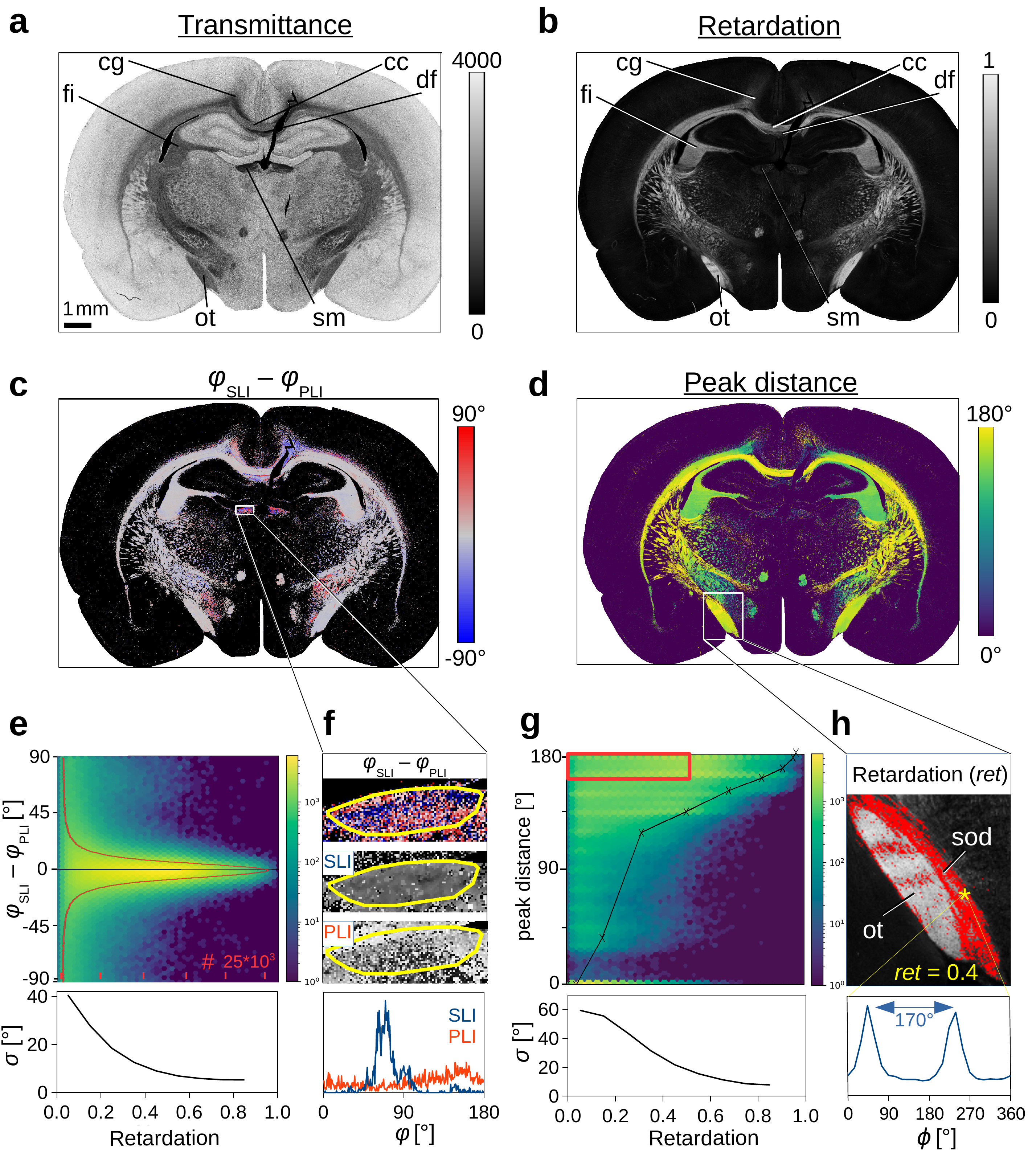}
	\caption{\textbf{Comparison of SLI and 3D-PLI measurements for a coronal rat brain section} (\cref{tab:table}\#78): \textbf{(a--b)} Transmittance and retardation images obtained from a 3D-PLI measurement, registered onto the SLI image stack. Anatomical regions are labeled for better reference (cg = cingulum, cc = corpus callosum, fi = fimbria, df = dorsal fornix, ot = optic tract, sm = stria medullaris). \textbf{(c)} Difference between in-plane fiber direction angles obtained from 3D-PLI and SLI (evaluated in regions with one or two prominent peaks with prominence $\geq 8\,\%$). \textbf{(d)} Distance between the peaks. \textbf{(e)} 2D histogram of all image pixels in (c) showing the difference between the in-plane fiber direction angles (SLI -- PLI) plotted against the retardation. Minimum values (zero) are shown in dark blue, maximum values in yellow (log-scale). The red curve in the 2D histogram shows the cumulated values (summed over all retardation values) for 256 bins between $\pm 90^{\circ}$. The graph below shows the standard deviation $\sigma$ of the angle difference plotted against the retardation (with a binning of 0.1). \textbf{(f)} Zoomed-in region of the stria medullaris. The histogram shows the fiber direction angles in the delineated region evaluated for SLI (blue) and 3D-PLI (red). \textbf{(g)} 2D histogram and standard deviation of the peak distances in (d) plotted against the retardation. The black curve in the 2D histogram shows peak distances of simulated line profiles for different out-of-plane fiber inclination angles (retardation values) obtained from \cref{fig:vervet-inclined-fibers}(d). \textbf{(h)} Zoomed-in region of the retardation image (optic tract). Values with a retardation $\leq 0.5$ and a peak distance $\geq 160^{\circ}$ (red rectangle in (g)) are marked in red. The graph shows the SLI profile evaluated in a region of $10 \times 10$ pixels (yellow asterisk, supraoptic decussation (sod) with retardation 0.4).}
	\label{fig:rat}
\end{figure}

To enable a direct comparison between the nerve fiber directions obtained from SLI and 3D-PLI, a 3D-PLI measurement of the same brain section was performed directly after the SLI measurement, and the images (average transmitted light intensities) were registered onto each other.
\Cref{fig:rat} shows the comparison between SLI and 3D-PLI for a coronal rat brain section. 
For the comparison, we considered only white matter regions with non-crossing nerve fibers, \ie\ regions with one or two prominent peaks in the SLI profiles (prominence $\geq 8\,\%$). 

Subfigures (c) and (e) show the differences between the in-plane fiber direction angles $\varphi$ obtained from SLI and 3D-PLI.\footnote{Note that $\varphi = 0^{\circ}$ is here defined along the positive x-axis, running counter-clockwise (while illumination angle $\phi = 0^{\circ}$ is defined along the positive y-axis, running clockwise).}
The histogram in (e) shows that the differences are distributed around zero with a full width half maximum of about $15^{\circ}$ (for about 766\,300 evaluated image pixels). The distribution of differences strongly depends on the out-of-plane inclination angle of the nerve fibers and, thus, on the retardation:
In regions with out-of-plane nerve fibers (\ie\ regions with low retardation), SLI and 3D-PLI yield very different direction angles (red and blue dots in (c)), \eg\ in the \textit{cingulum (cg)}, \textit{dorsal fornix (df)}, or \textit{stria medullaris (sm)}. The standard deviation is more than $20^{\circ}$ for regions with retardations $\leq 0.3$ and up to $40^{\circ}$ for regions with retardations $\leq 0.1$ (see lower graph in (e)).
In regions with in-plane nerve fibers (\ie\ regions with large retardation), SLI and 3D-PLI yield very similar direction angles (gray in (c)), \eg\ in the \textit{corpus callosum (cc)}, \textit{fimbria (fi)}, or \textit{optic tract (ot)}. The standard deviation is about $5^{\circ}$ or less for regions with retardations $\geq 0.7$. Taking into account that the error induced by the $15^{\circ}$-discretization of the SLI profiles is already $\pm 2.4^{\circ}$ (see \ref{sec:peak-positions}), the fiber direction angles from SLI correspond very well to those from 3D-PLI and can be considered as reliable. 

Subfigure (f) shows the direction angles of the SLI and 3D-PLI measurements in the stria medullaris, and the graph at the bottom shows their distribution in the delineated region. 
In the stria medullaris, a fiber tract is running steeply with respect to the coronal section plane. In such regions, 3D-PLI is prone to random in-plane direction angle distributions due to the low retardation signal (b).
While the 3D-PLI measurement (red curve) yields randomly distributed direction angles, the SLI measurement (blue curve) yields one dominant fiber direction around $\varphi_{\rm SLI} \approx 70^{\circ}$. This result is in accordance with the course of the fiber bundles in the stria medullaris which are expected to have a vertical but no lateral component in the coronal section plane, slightly declining on their rostral course. Hence, in some regions with out-of-plane nerve fibers, the SLI measurement patches the lack of reliability in 3D-PLI measurements and can be used as corrective. However, it should be noted that the SLI peaks merge for out-of-plane fibers and become smaller (as shown later in \cref{fig:vervet-inclined-fibers}(b)), which increases the risk of peaks not being detected and fiber directions determined from SLI profiles being shifted to wrong values.

\Cref{fig:rat}(d) and (g) show the distance between two peaks in the SLI profiles (evaluated in regions with one or two prominent peaks) in dependence on the retardation. A peak distance of $0^{\circ}$ means that only a single peak in the SLI profiles was detected. While regions with high retardation values (in-plane fibers) show always large peak distances between $160^{\circ}$ and $180^{\circ}$ as expected, regions with low retardation values (out-of-plane fibers or regions with less myelinated fibers) exhibit a broad distribution of peak distances from about $30^{\circ}$ to $180^{\circ}$. The standard deviation ranges from $7.7^{\circ}$ (at retardations above 0.8) to $59.4^{\circ}$ (at retardations below 0.1). 

In subfigure (h), regions with large peak distances ($160^{\circ}$--$180^{\circ}$) and small retardation values ($\leq 0.5$) are highlighted in red. Although the optic tract (in white) has much higher retardation values than the neighboring fiber bundle (supraoptic decussation, in red), the SLI profiles (here shown for a region with retardation 0.4) show similar peak distances ($170^{\circ}$--$180^{\circ}$) for both fiber bundles, suggesting that the two fiber bundles contain only moderately inclining nerve fibers. In fact, the supraoptic decussation follows mostly the course of the optic tract in this region \cite{paxinos2007}, but it contains less myelinated nerve fibers than the optic tract, which explains why we obtain lower retardation values. Without any correction, this would lead to an over-estimation of the out-of-plane nerve fiber inclinations in 3D-PLI. 

Hence, SLI yields not only a good reference for the fiber directions in regions with inclined nerve fibers, but also in other regions with low 3D-PLI signals like regions with lower myelination.


\enlargethispage{0.5cm}
\subsubsection{Inclined nerve fiber bundles}
\label{sec:whole_inclined}

\begin{figure}[htbp]
	\centering
	\includegraphics[width=0.9\textwidth]{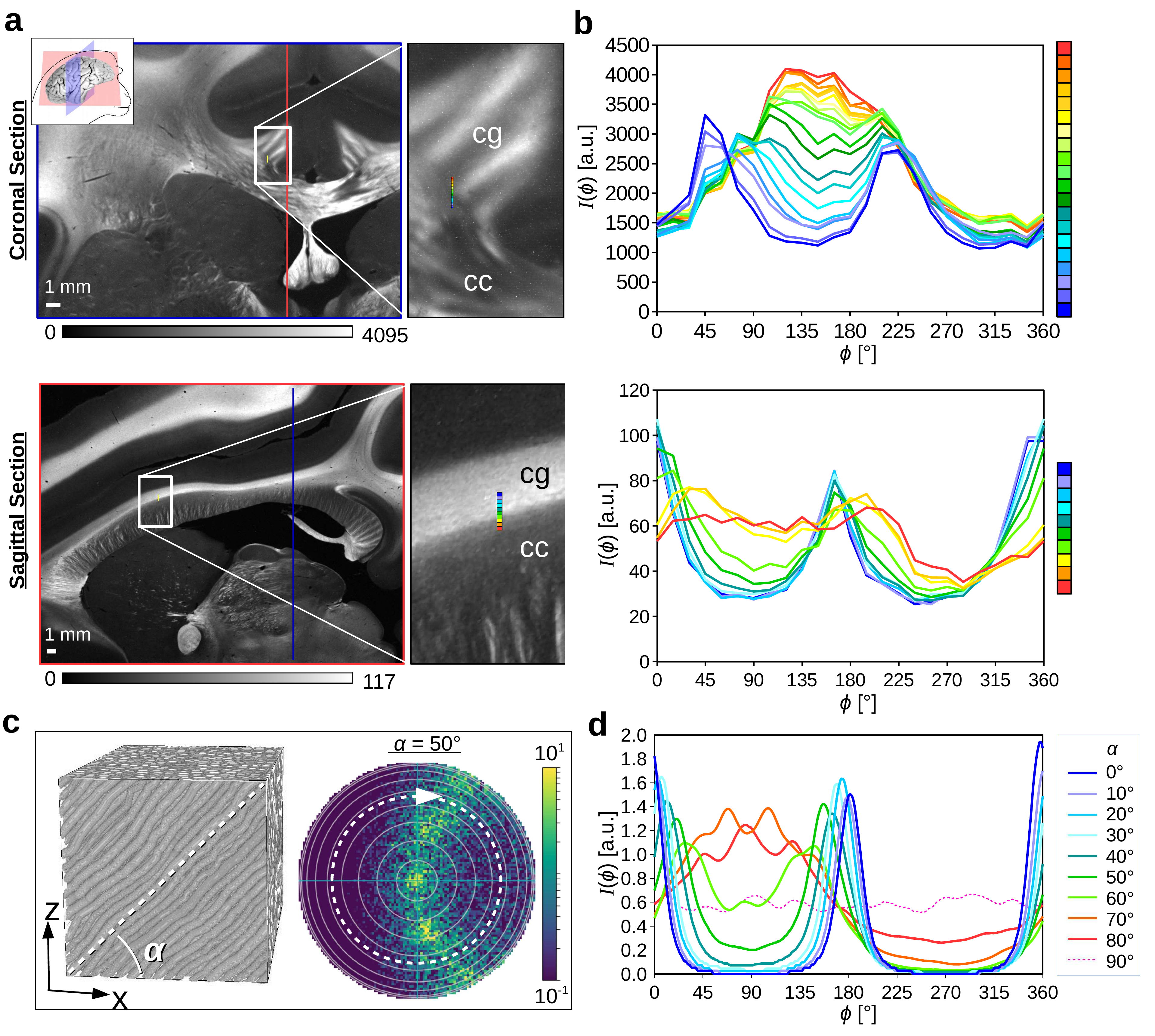}
	\caption{\textbf{Transition between in-plane and out-of-plane fibers.} \textbf{(a)} Coronal and sagittal section of a vervet brain (\cref{tab:table}\#161 and \#139). The images show the transmitted light intensity in the SLI measurement for an illumination from 12 o'clock ($\phi = 0^{\circ}$). The approximate location of the coronal (sagittal) section plane is marked in the other section plane by a blue (red) line. The zoomed-in areas show the transition between the corpus callosum (cc) and the cingulum (cg). \textbf{(b)} Cranio-caudal course of averaged SLI profiles for a chain of neighboring regions with $3 \times 3$ image pixels, evaluated along the rainbow-colored lines in the zoomed-in areas in (a). \textbf{(c)} Left: Artificial fiber bundle with out-of-plane inclination angle $\alpha$. Right: simulated scattering pattern for $\alpha = 50^{\circ}$. The images were adapted from \cite{menzel2020}, fig.\ 7(a), licensed with CC BY 4.0. \textbf{(d)} Simulated line profiles for different inclination angles $\alpha$. The line profiles were computed from the simulated (blurred) scattering patterns as described in \cref{sec:simulation} (average intensities evaluated along the white dashed circle in (c)).}
	\label{fig:vervet-inclined-fibers}
\end{figure}

To study how the SLI signal depends on the inclination angle of the fibers, we considered the transition zone between in-plane and out-of-plane fiber bundles in the corpus callosum (cc) and cingulum (cg) of a coronal and sagittal vervet brain section (see \cref{fig:vervet-inclined-fibers}). In the coronal (sagittal) section, the fibers in the corpus callosum are mostly oriented in-plane (out-of-plane). For fibers in the cingulum, it is the other way around.
To study the transition between in-plane and out-of-plane fiber bundles, the SLI profiles were evaluated for a chain of neighboring regions ($3 \times 3$ pixels, see rainbow-colored lines in (a)).
The blue/violet curves belong to regions with in-plane fiber bundles, the yellow/red curves to regions with out-of-plane fiber bundles, the green curves to the transition zone. In-plane fiber bundles (coronal: cc, sagittal: cg) show two distinct peaks lying $180^{\circ}$ apart; strongly inclined fiber bundles (coronal: cg, sagittal: cc) show one broad peak. With increasing inclination, the peak distance and the peak prominences decrease, while the peak widths increase. 
The same behavior was also observed for the coronal rat brain section (see \cref{fig:rat_supplement}).

The measured SLI profiles were compared to simulated line profiles (\cref{fig:vervet-inclined-fibers}(d)).
As observed for the SLI profiles (b), the peak distances in the simulated line profiles decrease, the peak widths increase, and the peak prominences decrease with increasing fiber inclination, until the two peaks merge into one broad peak for $\alpha \geq 80^{\circ}$. In \cref{fig:rat}(g), the peak distances of the simulated line profiles are plotted against the measured retardation of the rat brain section ($\vert\sin\delta\vert = \vert\sin(\arcsin({\rm ret}_{\rm max}) \, \cos^2\alpha)\vert$), where ${\rm ret}_{\rm max}$ is the maximum retardation value of the brain section, connecting the tissue retardation to the fiber inclination $\alpha$. 
The inclination dependence of the measured SLI profiles corresponds well\footnote{Due to less myelinated nerve fibers, the inclination angles are over-estimated (\ie\ shifted to lower retardation values) so that the simulated results set only a lower limit.} to the simulated results, validating again the benefit of the light scattering approach for elucidating the nerve fiber architecture in the brain.


\subsubsection{In-plane crossing nerve fibers}
\label{sec:whole_crossing}

\begin{figure}[htbp]
	\includegraphics[width=0.85\textwidth]{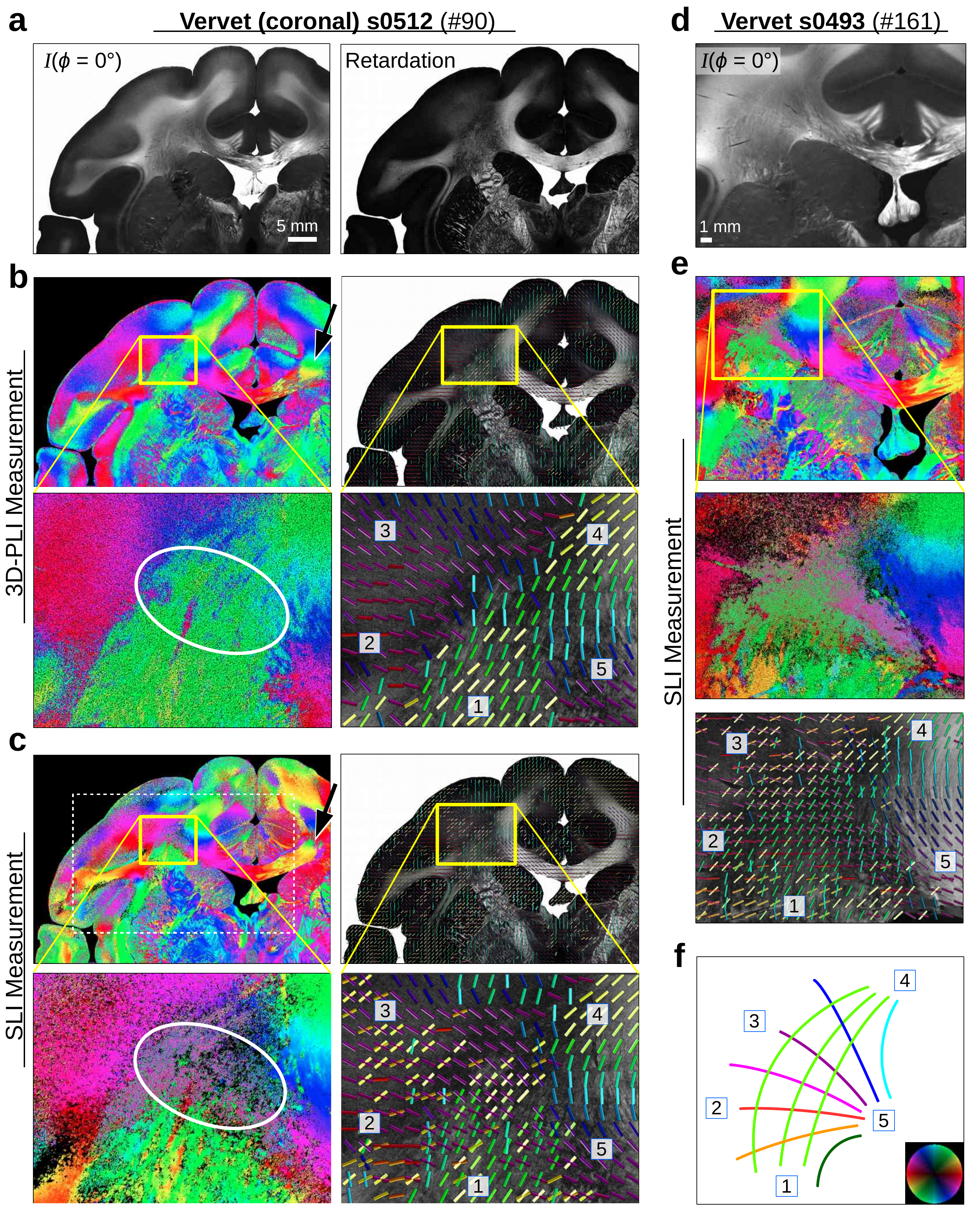}
	\caption{\textbf{Reconstructed nerve fiber directions of coronal vervet brain sections} (\cref{tab:table}\#90 and \#161): \textbf{(a)} Left: SLI image ($\phi = 0^{\circ}$) of vervet brain section no.\ 512. Right: retardation image obtained from a 3D-PLI measurement of the same section.
		\textbf{(b)} In-plane fiber directions obtained from the 3D-PLI measurement (px = 1.33\,\textmu m). \textbf{(c)} Fiber directions obtained from the SLI measurement (px = 13.7\,\um). \textbf{(d)} SLI image ($\phi = 0^{\circ}$) of vervet brain section no.\ 493 (px = 6.5\,\um). \textbf{(e)} Fiber directions obtained from the SLI measurement of the sample. The fiber directions are encoded in different colors (according to the color legend in (f)). The colored images show the fiber directions for each image pixel. The vector maps show the fiber direction for every $40^{\rm th}$ image pixel (b,c) and for every $50^{\rm th}$ image pixel (e). The SLI fiber directions were replaced by the median fiber direction of every $3 \times 3$ pixels beforehand. For better reference, the retardation image is shown in the background of the vector maps, respectively. The yellow rectangles mark the region of the corona radiata shown in the zoomed-in areas. \textbf{(f)} Sketch of crossing fiber pathways in the corona radiata, known from vervet brain atlases \cite{woods2011}. The start/end points of the pathways are labeled by numbers (also shown in the zoomed-in areas of the vector maps).}
	\label{fig:vervet}
\end{figure}

To demonstrate the potential of SLI to reconstruct crossing nerve fiber directions in whole brain tissue samples, coronal vervet brain sections were measured both with SLI and 3D-PLI, and the (registered) fiber directions were compared to each other at different levels of detail (see \cref{fig:vervet}). A special focus was placed on the corona radiata region (yellow rectangles), representing a region with complex nerve fiber crossings (cf.\ sketch in (f)).

The pathways of the crossing fibers in the corona radiata (which are already visible in the structure of the retardation image from 3D-PLI) become clearly visible in the SLI direction images. For better reference, the retardation image is shown in the background of the vector maps. As expected, regions with small retardation values correspond to regions with crossing nerve fibers, where SLI yields additional information. 
The 2D fiber directions reconstructed from SLI correspond to the course of the fiber pathways shown in (f): In addition to the course of the corpus callosum (5)--(4), which is already visible in the 3D-PLI vector map (b), the vector map in (c) also shows the fiber bundle starting at (5) and fanning out to (2) and (3) in red/magenta, as well as the fiber pathways between (2)--(3) and (1)--(4) in green/yellow. The same pattern of crossing fibers is also visible in the SLI vector map of the neighboring vervet brain section (e).


\section{Discussion}

For building a detailed network model of the brain, a correct reconstruction of crossing nerve fiber pathways is crucial.
Diffusion magnetic resonance imaging (dMRI) allows to measure crossing nerve fiber orientations, but its resolution in post-mortem brains is limited to a few hundred micrometers, hence limiting its application for reconstructing individual nerve fiber tracts.
In principle, 3D Polarized Light Imaging (3D-PLI) can be used to reconstruct the orientations of crossing nerve fibers:
By taking directional information of neighboring image pixels into account, the orientations of crossing nerve fibers can be estimated \cite{axer2016}. 
However, the resolution is fundamentally limited because a ``super-pixel'' of several pixels (typically $10 \times 10$ pixels or more) needs to be formed.
Furthermore, 3D-PLI often yields erroneous (intermediate) nerve fiber orientations in regions with crossing nerve fibers
(cf.\ \cref{fig:chiasm_SLI_vs_PLI}(a)(ii) and (iii) in cyan) so that the super-pixel also contains erroneous fiber orientations.
To extract structural information such as (crossing) fiber directions from microscopic image data, structure tensor analysis (STA) is widely used \cite{khan2015,wang2015}. However, STA also comprises information from several neighboring image pixels, which limits the resolution, and it does not yield reliable fiber directions in bulk tissue without recognizable structures.

Another possibility to reveal (crossing) nerve fiber directions is to study the distribution of scattered light (scattering pattern) behind the sample.
Scatterometry allows to measure the full\footnote{The maximum scattering angle is limited by the numerical aperture of the imaging system (here: $\theta = \arcsin(0.4) \approx 23^{\circ}$).} scattering pattern, but only for brain regions $\geq 100$\,\textmu m \cite{menzel2020-BOEx}. In Scattered Light Imaging (SLI), we measure a limited number of scattering angles with a reverse setup, but for whole brain tissue samples at once (with a field-of-view of several centimeters)
and with an in-plane spatial resolution that is only limited by the pixel size in object space (currently: $\geq 6.5$\,\um). Another advantage is that SLI can be performed in a relatively short amount of time (without rasterizing), using simply a standard LED panel and a camera.

\subsection{Prototypic SLI measurements}

The measurements were performed with a simple and cheap setup, 
allowing other laboratories to immediately use SLI to study crossing nerve fibers in the brain. 
The prototypic setup turned out to be sufficient to validate our technique and to develop an automated evaluation method. 
As the measurement was performed manually with a physical mask, the measurement of many illumination angles is quite labor-intensive. Therefore, we chose $15^{\circ}$-steps as compromise between measurement time and accuracy. The developed correction procedure (\ref{sec:peak-positions}) reduces the discretization error to $\pm 2.4^{\circ}$. 

While the fiber directions in 3D-PLI can be determined from a sinusoidal fit (requiring only a small number of data points), there exists no simple fit function for SLI profiles. The interpretation of the signals depends on the number of detected peaks (cf.\ \cref{fig:scattering-line-profiles}(b)). By considering only prominent peaks, the number of correctly assigned regions was notably increased (see \ref{sec:prominence-determination}).

Even if some existing fiber directions are not determined, we could show that the assigned fiber direction angles can safely be assumed as correct.
Therefore, SLI can be used as cross-reference for other imaging methods. Missing fiber directions can be estimated from neighboring pixels. As a consequence, the effective resolution is reduced, but the achievable in-plane resolution (\eg\ $13\,\um$ for $2\times2$ pixels with a size of 6.5\,\textmu m) is still much higher than for dMRI.

While there exists a clear correlation between the peak positions of the SLI profiles and the fiber direction, the characteristics of the SLI profiles do not provide a reliable estimate for the distribution of fiber orientations. The peak width, for example, is expected to increase for broadly distributed fibers, but also increases with increasing fiber inclination angle (cf.\ \cref{fig:vervet-inclined-fibers}) or diameter. A clear distinction between these influences is very challenging and left for future work. For the same reason, it is currently not possible to provide a pixel-wise estimate for the confidence of the determined fiber directions. However, our studies suggest that the 2D fiber directions obtained from SLI are reliable and -- in contrast to 3D-PLI, where the birefringence signal decreases significantly with increasing inclination angle -- this also holds for strongly inclined (parallel) fibers (cf.\ \cref{fig:rat}).

\subsection{Validation of SLI}

To validate SLI, we used a combination of alternative imaging methods, tissue phantoms, simulations, and brain regions with known fiber architectures.

For in-plane nerve fibers, we compared results from SLI to results obtained from 3D-PLI and scatterometry for a large range of data points (\cref{sec:SLI-scatterometry-comparison,sec:whole_non-crossing}).
We found that the differences in fiber direction are small with a standard deviation of $\pm 8.8^{\circ}$ (scatterometry) and $\pm 5^{\circ}$ (3D-PLI, for regions with mostly non-crossing in-plane nerve fibers with retardations above 0.7).
Taking into account the different measurement setups (different resolutions, light sources, measurement times) and a discretization error of $\pm 2.4^{\circ}$ for the SLI profiles, this is a very good correspondence.
Also, the reconstructed directions of (crossing) nerve fibers correspond to anatomically known fiber pathways and to visible structures in the high-resolution 3D-PLI images (\cref{fig:vervet}).

As model system, we used two/three crossing sections of optic tracts (\cref{sec:chiasm}). 
Producing a sample with more than three crossing sections is not feasible: The sections would need to be very thin ($< 30$\,\textmu m) making the preparation of the sections and the arrangement under defined crossing angles difficult.
Therefore, the SLI fiber directions were only validated for up to three crossing fiber bundles. As all our validation approaches were consistent, our findings suggest that SLI allows to reconstruct more than three crossing nerve fibers, provided that the minimum crossing angle is larger than the resolution limit.

For out-of-plane nerve fibers, a simple model system is not available:
Cutting \eg\ the optic tract under oblique angles would not yield a sample with defined inclination angles because the tissue would be distorted when being placed between the glass slides. 
Tilting the whole sample (section of optic tract) would yield a different result because the fibers are not cut diagonally and would show a different scattering behavior than out-of-plane fibers in real tissue samples. 
For the same reason, a model system of inclined crossing nerve fibers is not available (\eg\ crossing sections of optic tracts with different tilting angles).

A comparison to scatterometry or 3D-PLI is also challenging because out-of-plane nerve fibers yield non-symmetric scattering patterns (leading to different scatterometry and SLI profiles, cf.\ \cref{sec:SLI-scatterometry-comparison}), and the fiber inclinations derived from 3D-PLI depend on other factors such as the myelination of the fibers.	Instead, the SLI profiles of out-of-plane nerve fiber bundles (anatomically known regions with inclined fibers) were compared to simulated\footnote{The simulations were performed for nerve fiber diameters between 1.0--1.6\,\um. Previous simulation studies (\cite{MMenzel}, Fig.\ 11.14) suggest that nerve fibers with up to five times larger diameters yield stronger and broader scattering peaks, but the peaks still occur at the same positions. Hence, the out-of-plane effects estimated by the simulations remain valid.} line profiles (\cref{sec:whole_inclined}). The results show that the SLI profiles are compatible with the simulated profiles. However, as the exact fiber inclinations in the measured brain section are not known, the comparison is only qualitative and cannot be used to determine the exact relationship between peak distance and fiber inclination.

For inclined crossing fibers, it is hard to find a corresponding region in real brain tissue with defined inclination angle so that an experimental validation is not possible. Simulations of two crossing fiber bundles with different inclination angles (see \cref{fig:cross_incl}) suggest that, in principle, it might be possible to identify inclined crossing fibers on behalf of their SLI profiles. However, as other nerve fiber constellations might yield similar scattering patterns, this is very challenging and left for future studies.

\subsection{Combination of SLI and 3D-PLI}
\label{sec:discussion}

The SLI measurements do not only provide additional information in regions with in-plane crossing nerve fibers, but also yield a good reference and corrective for fiber directions in other brain regions with small 3D-PLI signals. 
In some regions with steep out-of-plane nerve fibers, the 2D fiber directions derived from SLI were shown to be more reliable than those obtained from 3D-PLI (see \cref{fig:rat}(f)).
In some regions with less myelinated nerve fibers (supraoptic decussation in \cref{fig:rat}(h)), we could show that the SLI measurement also serves as sanity check for the fiber inclinations determined by 3D-PLI, which need to be corrected \eg\ by transmittance weighting.\footnote{Due to the high scattering coefficient of myelin, highly myelinated regions appear darker than less myelinated regions in the transmittance image. Thus, the transmittance image can be used to correct for regions with less myelinated nerve fibers.}
When using a tiltable specimen stage in 3D-PLI \cite{wiese2014,schmitz2018}, the fiber inclinations could be determined more reliably, also in regions with low fiber densities. However, tilting is not (yet) in use for microscopic resolutions < 10\,\um.

Due to the similar setup (illuminating unstained histological brain sections and measuring the transmitted light), the SLI measurement can be easily combined with a 3D-PLI measurement (by using removable polarizing filters in the 3D-PLI setup), which significantly improves the high-resolution 3D-\hskip0pt reconstruction of the nerve fibers: As SLI and 3D-PLI have different sources of error, SLI can serve as cross-validation for 3D-PLI, also when using a tiltable specimen stage. When performing SLI and 3D-PLI measurements on several consecutive brain sections, registering the resulting images onto each other (as described in \cref{sec:registration}), and combining the 2D vector field from SLI (containing up to three crossing fiber directions in each image pixel) with the 3D vector field from 3D-PLI (containing one three-dimensional fiber orientation for each image pixel), streamline-based tractography can be performed on the three-dimensional brain volume \cite{schubert2018}.

A combined measurement of SLI and 3D-PLI can also help to estimate the myelin content of brain tissue: The transmittance images obtained from 3D-PLI appear darker in highly myelinated regions, but also in regions with out-of-plane nerve fibers (as shown by \cite{menzel2020}). An additional SLI measurement can help to distinguish highly myelinated regions with in-plane crossing nerve fibers from less myelinated regions with out-of-plane nerve fibers, which both yield small birefringence signals and cannot be distinguished by 3D-PLI. This allows to use the transmittance as reliable estimate for the myelin content in the investigated brain section.

When performing a combined measurement of SLI and 3D-PLI on the same tissue sample, it should be noted that the embedding time has a different impact: For SLI, a long-term study (\ref{sec:long-term}) has shown that a longer time between tissue embedding and measurement increases the signal-to-noise ratio in regions with crossing nerve fibers (due to reduced higher-order scattering). For 3D-PLI, on the other hand, the long embedding time might cause problems because the tissue becomes more transparent so that the transmittance cannot be used anymore to correct the determined fiber inclinations.

\subsection{Limitations of SLI}

The discretization of the SLI profiles sets a lower limit to the accuracy of the determined peak positions (and thus the derived fiber directions) and to the minimum resolvable crossing angle between the nerve fibers. The accuracy can be easily improved by increasing the number of illumination angles.
The minimum resolvable crossing angle, however, will still be limited due to the finite widths of the scattering peaks. Scatterometry measurements \cite{menzel2020-BOEx} suggest that nerve fibers in typical brain tissue cannot be properly distinguished if the crossing angle lies below $25^{\circ}$.

The in-plane resolution of SLI is limited by diffraction and the available objective lens.
To avoid artifacts, the pixel size should still be larger than the average nerve fiber diameter. For 3D-PLI measurements, a pixel size of 1.33\,\textmu m turned out to be a good compromise (expected to be similar for SLI).

While the in-plane resolution of SLI can be in the order of a micron, the effective anatomical resolution is limited by the thickness of the brain section ($\geq 20$\,\um, to ensure sections are not too fragile). The SLI profiles reflect the combined scattering effect of all nerve fibers contained within one tissue voxel; small variations between fiber orientations within one voxel cannot be detected.

The software \texttt{SLIX} enables a reliable reconstruction of in-plane (crossing) nerve fibers. To account for slightly inclined nerve fibers, a range of pair-wise peak distances ($180^{\circ} \pm 35^{\circ}$) is considered. However, as the exact relationship between peak distance and fiber inclination is not known, the software does not provide a reconstruction of out-of-plane angles or a solution for inclined crossing nerve fibers. This will be left for future studies.

\subsection{Perspectives for further improvements}

One major source of error is the incorrect interpretation of peaks in SLI profiles.
Future studies should focus on how to combine different parameter maps (cf.\ \cref{fig:chiasm-parametermaps}) to improve the reconstruction of nerve fiber directions also in regions with less prominent peaks, \eg\ by including the peak prominence as a measure for the signal-to-noise ratio.
Future versions of the \texttt{SLIX} software should also include correction procedures to account for non-determined fiber directions. Reconstructing missing information from an image is a well-studied problem in computer science and mathematics, and much simpler than correcting erroneous information (see \cite{gross2011} and references therein).

With the developed software, it is possible to perform a systematic evaluation of anatomically known brain structures, to study if other fiber constellations (\eg\ inclined crossing fibers) can be reconstructed with SLI and/or by combining SLI with 3D-PLI measurements.

To enable more precise measurements, the SLI measurement should be fully automated so that a larger number of scattering angles can be measured.
Using a display with individually controllable LEDs would allow the measurement of full scattering patterns (by performing individual measurements for all LEDs in the display). Future studies should investigate whether the measurement of additional scattering angles yields important information about the substructure of the tissue and allows to reconstruct \eg\ inclined crossing fibers.

The minimum pixel size in the SLI measurement ($6.5\,\um$) can still be improved by using a different objective lens.
To avoid artifacts caused by image registration and different resolutions, the SLI and 3D-PLI measurements should be integrated in the same setup. The results obtained from the SLI measurements can be used to validate and correct the nerve fiber orientations reconstructed by 3D-PLI and notably improve the nerve fiber tractography.
When studying inclined crossing fibers, a combined measurement of SLI and 3D-PLI would also be helpful: While the SLI profiles might be ambiguous, low transmittance and low retardation values indicate inclined crossing nerve fibers, as proposed in \cite{menzel2020}.


\section{Conclusions}

The detailed reconstruction of crossing nerve fibers in the brain poses a major challenge for many neuroimaging techniques.
Here, we introduced Scattered Light Imaging (SLI), a neuroimaging technique that allows to reconstruct the individual directions of (crossing) nerve fibers with micrometer in-plane resolution over a field of view of several centimeters. 
In measurements of different species (human, vervet, rat), we verified that SLI yields reliable 2D fiber directions, in particular in regions with crossing nerve fibers.
SLI can be easily combined with 3D-Polarized Light Imaging (3D-PLI). This allows to benefit both from the high-precision three-dimensional nerve fiber orientations from 3D-PLI and the full structural information from SLI, greatly enhancing the reconstruction of the highly complex nerve fiber architecture in the brain.
In conclusion, SLI is a major step towards a full reconstruction of the brain's nerve fiber network with micrometer resolution, especially in regions with crossing nerve fibers.


\section*{Data and code availability statement}
All images used in this study (original image stacks and SLI parameter maps, as well as the transmittance, direction, and retardation images from 3D-PLI) are available in original resolution on the EBRAINS data repository \cite{EBRAINS}. The code used in this study has been published as open-source software \texttt{SLIX} (Scattered Light Imaging ToolboX) and is available on GitHub (\url{https://github.com/3d-pli/SLIX}).
For this work, \texttt{SLIX} v1.2.2 was used.

\section*{Declaration of competing interest}
None.

\section*{CRediT authorship contribution statement}
\textbf{Miriam Menzel:} Conceptualization, Methodology, Investigation, Formal analysis, Validation, Visualization, Software, Supervision, Writing - original draft, review \& editing.
\textbf{Jan A.\ Reuter:} Software, Formal analysis, Validation, Visualization, Writing - review \& editing.
\textbf{David Gräßel:} Methodology, Investigation, Writing - review \& editing.
\textbf{Mike Huwer:} Investigation.
\textbf{Philipp Schlömer:} Methodology, Investigation, Supervision.
\textbf{Katrin Amunts:} Funding acquisition, Supervision, Writing - review \& editing.
\textbf{Markus Axer:} Project administration, Supervision, Writing - review \& editing.


\section*{Acknowledgements}
We thank Markus Cremer, Patrick Nysten, and Steffen Werner for the preparation of the histological brain sections, Martin Schober for assistance with image registration, and Nicole Schubert and Felix Matuschke for assistance with the visualization of nerve fiber directions. 
Furthermore, we thank Karl Zilles and Roger Woods for collaboration in the vervet brain project (National Institutes of Health under Grant Agreements No.\ R01MH092311 and No.\ 5P40OD010965), and Roxana Kooijmans and the Netherlands Brain Bank for providing the human optic chiasm.
This project has received funding from the European Union's Horizon 2020 Framework Programme for Research and Innovation under the Specific Grant Agreement No.\ 785907 and 945539 (``Human Brain Project'' SGA2 and SGA3).
M.M.\ received funding from the Helmholtz Association port-folio theme ``Supercomputing and Modeling for the Human Brain'' and the Helmholtz Doctoral Prize 2019.
We gratefully acknowledge the computing time granted through JARA-HPC on the supercomputer JURECA at Forschungszentrum Jülich \cite{jureca2}.

\cleardoublepage
\appendix


\section{Optimum prominence value for peak detection}
\label{sec:prominence-determination}
\begin{figure}[h!]
	\centering
	\includegraphics[width=0.5\textwidth]{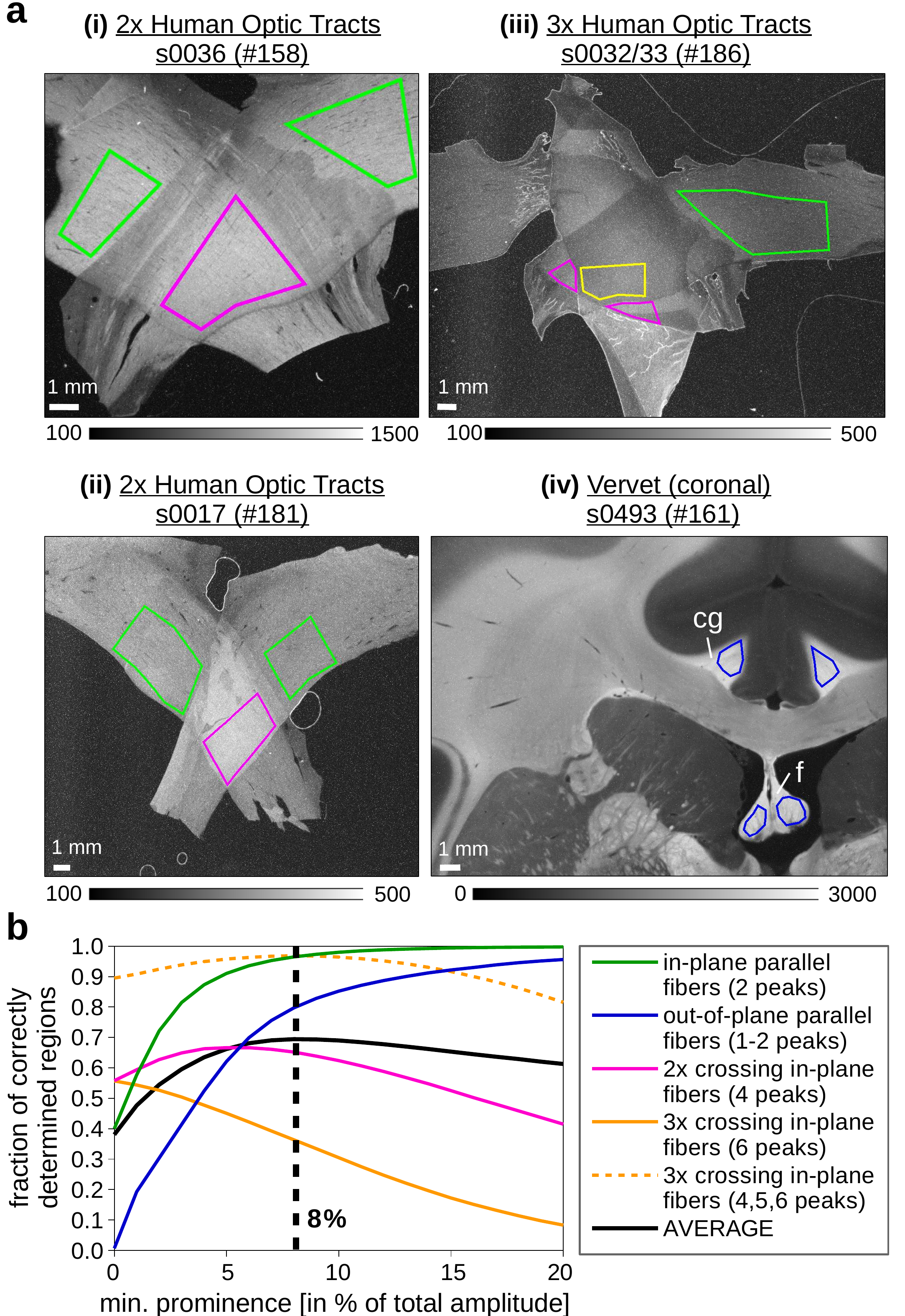}
	\caption{\textbf{Determination of the optimum prominence value for peak detection.} \textbf{(a)} Average transmitted light intensity in an SLI measurement of two and three crossing sections of optic tracts ((i)-(iii)) and a coronal vervet brain section (iv). Regions used for evaluation are surrounded by colored lines: in-plane parallel fibers (green), two crossing fiber bundles/layers (magenta), three crossing fiber bundles/layers (yellow), and out-of-plane fibers (blue). cg = cingulum; f = fornix. \textbf{(b)} Fraction of correctly determined regions plotted against the minimum prominence value used for peak detection evaluated for the different types of regions shown in (a). The black curve shows the average of all curves.}
	\label{fig:prominence}
\end{figure}

The peaks in the measured SLI profiles are not always related to nerve fiber directions in the measured tissue sample. They can also be caused by noise or tissue irregularities. To avoid misinterpretations, not all peaks in the SLI profiles were considered for evaluation, but only peaks with a certain prominence (cf.\ \cref{fig:scattering-line-profiles}(b), in red). To identify the optimum prominence value, we compared the expected number of peaks to the number of peaks computed in regions with known underlying fiber architecture (see \cref{fig:prominence}(a)): regions with in-plane parallel fibers (green) where two peaks are expected, regions with out-of-plane parallel fibers (blue) where one or two peaks are expected, regions with two crossing in-plane fiber bundles (magenta) where four peaks are expected, and regions with three crossing in-plane fiber bundles (orange) where six peaks are expected. To minimize errors, we considered only regions with mostly homogeneous and well-known fiber architectures: in-plane (non-)crossing fibers in manually crossed sections of optic tracts with different crossing angles, and out-of-plane fibers in the cingulum/fornix of a coronal vervet monkey brain section. The SLI measurements were performed as described in \cref{sec:SLI} with a mask with rectangular holes and $15^{\circ}$-steps.
For each region marked in \cref{fig:prominence}(a), the SLI profiles were computed for each image pixel and the peaks in the SLI profiles were determined as described in \cref{sec:evaluation-line-profiles} for different minimum prominence values: \{0\,\%, 1\,\%, $\dots$, 20\,\%\} of the total signal amplitude. The computed number of peaks was then compared to the expected number of peaks for each image pixel in the different types of regions. 

\Cref{fig:prominence}(b) shows the fraction of correctly determined regions (\ie\ the number of pixels where the computed peak number corresponds to the expected peak number, divided by the total number of pixels in the considered type of region) for different minimum prominence values.
In regions with parallel in-/out-of-plane fibers (green/blue curves) where one or two large peaks are expected, the fraction of correctly determined regions becomes largest when only peaks with larger prominence values are considered: when only peaks with a prominence values $\geq 20$\,\% are considered, 99.7\,\% (95.6\,\%) of the regions with in-plane (out-of-plane) fibers are correctly determined.
In regions with three crossing fiber bundles/layers (orange curve) where the distance between the resulting peaks is small, the fraction of correctly determined regions becomes largest (55.7\,\%) when all peaks (\ie\ with prominences $\geq 0$\,\%) are considered. In regions with two crossing fiber bundles/layers (magenta curve) where the distance between the resulting peaks is larger, the fraction of correctly determined regions becomes largest (66.7\,\%) for a minimum prominence value of 5\,\%: For smaller values, too many (\ie\ also erroneous) peaks are considered. For larger values, not enough (less than four) peaks are considered. 

To find a good compromise between regions with different fiber constellations, the curves for in-plane (crossing) and out-of-plane fibers were averaged with equal weight (see black curve in \cref{fig:prominence}(b)). For a minimum prominence value of 8\,\%, the fraction of correctly determined regions becomes largest (69.4\,\%). For this value, 96.5\,\% of in-plane parallel fibers, 79.8\,\% of out-of-plane parallel fibers, 65.2\,\% of two crossing fiber bundles, and 36.3\,\% of three crossing fiber bundles are correctly determined. The low value for the three crossing fiber bundles is due to the $15^{\circ}$-discretization of the SLI profiles and the relatively small amplitude of the resulting peaks (cf.\ \cref{fig:chiasm-parallel-cross-lineprofiles} on the right). Most of these regions (96.7\,\%) are still determined as crossing fibers (4, 5, or 6 peaks, see orange dotted curve).
Therefore, when evaluating SLI profiles, we used a minimum prominence of 8\,\% for the peak detection, as described in  \cref{sec:evaluation-line-profiles}.


\section{Correction of determined peak positions}
\label{sec:peak-positions}

\begin{figure}[h!]
	\centering
	\includegraphics[width=0.7\textwidth]{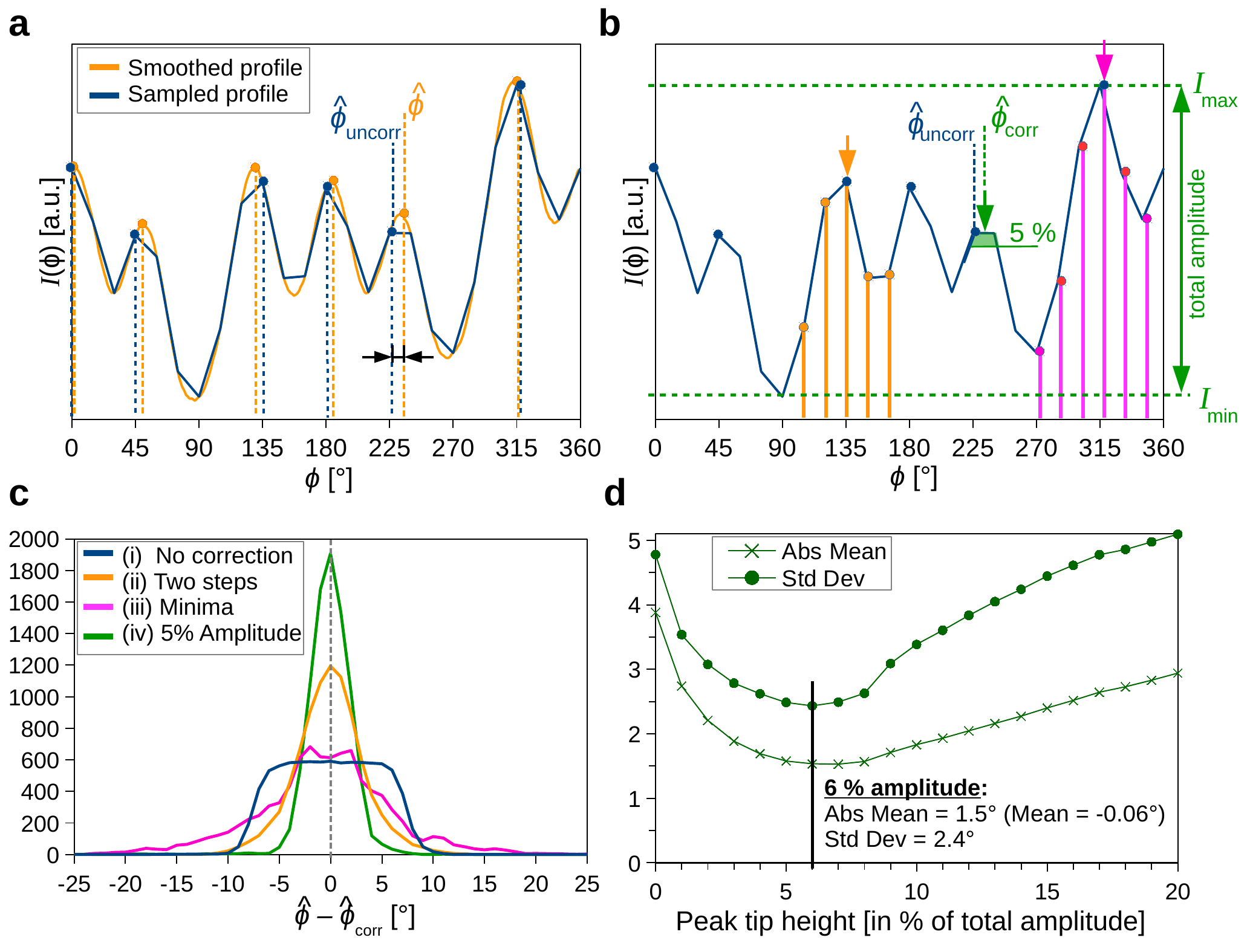}
	\caption{\textbf{Correction of determined peak positions in the SLI profiles.} \textbf{(a)} The orange curve shows the smoothed line profile (scatterometry profile) for three crossing sections of optic tracts (the measured region is marked in \cref{fig:samples} by a yellow arrow). The blue curve shows the line profile sampled in steps of $15^{\circ}$ (starting at $\phi = 0^{\circ}$). Because of the sampling, the determined peak positions $\hat{\phi}_{\text{uncorr}}$ (vertical blue dashed lines) do not exactly correspond to the original peak positions $\hat{\phi}$ in the smoothed line profile (vertical orange dashed lines). \textbf{(b)} The determined peak positions $\hat{\phi}_{\text{uncorr}}$ (blue circles) were modified with different correction methods, yielding corrected peak positions $\hat{\phi}_{\text{corr}}$ (colored arrows): Centroid over $\pm 2$ steps (orange), centroid between two minima (pink), centroid of peak tip (green) for a height of 5\,\% of the total amplitude. \textbf{(c)} The histograms show the difference between the original peak positions in the smoothed line profiles ($\hat{\phi}$) and the corrected peak positions in the sampled line profiles ($\hat{\phi}_{\text{corr}}$) for the different methods. In total, $15 \times 232$ sampled line profiles were compared to 232 smoothed line profiles (scatterometry profiles) of 9 different samples (see all circles in \cref{fig:samples}). \textbf{(d)} Absolute mean and standard deviation of the histogram ($\hat{\phi} - \hat{\phi}_{\text{corr}}$), where the peak positions were corrected by computing the centroid of the peak tips at different heights ($\{0\,\%,1\,\%,\dots,20\,\%\}$ of the total signal amplitude). For 6\,\%, the standard deviation becomes minimal ($2.4^{\circ}$).}
	\label{fig:PeakPos-Correction}
\end{figure}

During an SLI measurement, the hole of the mask is rotated in steps of $\Delta\phi = 15^{\circ}$ or $22.5^{\circ}$ (see \cref{sec:SLI}), yielding a discretized SLI profile $I(\phi)$ for each image pixel. The peak positions in the discretized SLI profiles usually do not correspond to the underlying nerve fiber directions in the measured tissue sample (cf.\ \cref{fig:PeakPos-Correction}(a)). Assuming that the scattering peaks generated by nerve fibers are mostly symmetric and do not influence each other, we can use the geometric centers (\textit{centroids}) of the peaks as an estimate for the ``real'' peak positions of the ideal, non-discretized line profile:
\begin{align}
	\hat{\phi}_{\text{corr}} = \frac{\sum\limits_{\phi = \hat{\phi}_{\text{uncorr}}-\phi_L}^{\hat{\phi}_{\text{uncorr}} +\phi_R} I(\phi) \cdot \phi}{\sum\limits_{\phi = \hat{\phi}_{\text{uncorr}}-\phi_L}^{\hat{\phi}_{\text{uncorr}} +\phi_R} I(\phi)} \,\,.
	\label{eq:centroid}
\end{align}
In this notation, $\hat{\phi}_{\text{uncorr}}$ is the peak position in the discretized line profile, $\hat{\phi}_{\text{corr}}$ is the estimated peak position of the non-discretized line profile, $\phi_L$ and $\phi_R$ are the left and right limits used for the centroid calculation, and $I(\phi)$ is the measured light intensity for the illumination angle $\phi$.

To find the optimum values for the left and right limits, we compared different methods (see \cref{fig:PeakPos-Correction}(b)):
\begin{enumerate}[(i)]
	\item \textbf{No correction:} The peak positions were determined as described in \cref{sec:evaluation-line-profiles}, without any correction (blue circles).
	\item \textbf{Centroid over $\pm 2$ steps}: The centroids of the determined peaks (orange arrow) were computed with eq.\ (\ref{eq:centroid}) over two steps of discretization to the left/right ($\phi_L = \phi_R = 30^{\circ}, \Delta \phi = 15^{\circ}$).
	\item \textbf{Centroid between two minima:} The centroids of the determined peaks (pink arrow) were computed with eq.\ (\ref{eq:centroid}), using the nearest left and right minima (with prominence $\geq 8$\,\%) as left and right limits. When there was no such minimum on one side, a distance of two steps ($30^{\circ}$) was chosen as limit.
	\item \textbf{Centroid of peak tip:}  All $15^{\circ}$-steps in the discretized line profile were divided into 100 equidistant points and linearly interpolated between the points, yielding steps of $\Delta\phi = 0.15^{\circ}$. From each determined peak position, the points in the line profiles were followed to the left and the right until the distance to the peak becomes larger than two steps ($30^{\circ}$), a minimum (with prominence $\geq 8$\,\%) is reached, or the intensity value at one point reaches a value below \{0\,\%, 1\,\%, \dots, 20\,\%\} of the total amplitude ($I_{\text{max}} - I_{\text{min}}$) of the line profile. The centroid of the peak (green arrow) was computed with eq.\ (\ref{eq:centroid}), using these limits (here: for a peak tip height of 5\,\% of the total signal amplitude).
\end{enumerate}
As we only want to correct for discretization artifacts, the peak positions were shifted by maximally one step size: $\vert\hat{\phi}_{\text{corr}} - \hat{\phi}_{\text{uncorr}}\vert \leq 15^{\circ}$.

To test the different correction methods, we used 232 non-discretized scatterometry profiles (obtained from 9 different brain tissue samples, see all circles in \cref{fig:samples}) as model systems.
To obtain a discretized line profile for the comparison, the scatterometry profiles were sampled in steps of $15^{\circ}$  (see \cref{fig:PeakPos-Correction}(a)).
In real brain tissue, the fiber orientations are not known with $< 15^{\circ}$ precision, \ie\ each fiber direction (peak position) is equally probable within $15^{\circ}$. Although the scatterometry profiles belong to different fiber configurations, the underlying fiber orientations are not uniformly distributed within $15^{\circ}$. To generate an unbiased, uniformly distributed data set reflecting all possible fiber orientations (samplings) that might occur in an SLI measurement, 15 different starting points \{$0^{\circ}$, $1^{\circ}$, $\dots$, $14^{\circ}$\} were used for the sampling, yielding 15 discretized line profiles for each scatterometry profile. 
For all $15 \times 232$ discretized line profiles, the corrected peak positions $\hat{\phi}_{\text{corr}}$ were computed, using the four correction methods described above (colored arrows in \cref{fig:PeakPos-Correction}(b)), and compared to the original peak positions $\hat{\phi}$ of the scatterometry profile (orange circles in \cref{fig:PeakPos-Correction}(a)). To ensure that only positions belonging to the same peak are compared to each other, we considered only line profiles with the same number of peaks. In total, 8759 peaks in 3301 discretized line profiles could be used for comparison. For 4.7\,\% (0.4\,\%) of the line profiles, one (two) peaks were missing after sampling.

\enlargethispage{0.5cm}
The histograms in \cref{fig:PeakPos-Correction}(c) show the difference between original and corrected peak positions for the different correction methods. 
All curves are centered around $0^{\circ}$ (the means lie between $-0.01^{\circ}$ and $-0.65^{\circ}$), but they differ in terms of absolute mean and standard deviation: (i) Without any correction (blue curve), the differences are almost uniformly distributed between $\pm 5^{\circ}$ (absolute mean: $3.9^{\circ}$, standard deviation: $4.8^{\circ}$). As every peak is sampled in $15^{\circ}$-steps with 15 different offsets, the resulting peak positions of the sampled line profiles are expected to be uniformly distributed around the original peak position within $\pm (15^{\circ}/2)$ in case of symmetric peaks. In case of asymmetric peaks, the differences can be larger. 
(ii) When correcting the peak positions by computing the centroid between neighboring minima (pink curve), the kurtosis increases, but the corrected peak positions can still differ a lot from the original peak positions (absolute mean: $4.9^{\circ}$, standard deviation: $6.5^{\circ}$). 
(iii) When computing the centroid over $\pm 2$ steps (orange curve), the distribution becomes much narrower and resembles a normal distribution (absolute mean: $2.6^{\circ}$, standard deviation: $3.7^{\circ}$). 
(iv) When computing the centroid of the peak tip with a height of 5\,\% of the total signal amplitude (green curve, cf.\ \cref{fig:PeakPos-Correction}(b) in green), the distribution has an even smaller standard deviation (absolute mean: $1.6^{\circ}$, standard deviation: $2.5^{\circ}$).

\Cref{fig:PeakPos-Correction}(d) shows the absolute mean and standard deviation for method (iv) for different peak tip heights ($\{0\,\%, 1\,\%, \dots, 20\,\%\}$  of the total amplitude). If the height becomes too small, there are not enough points of the sampled line profile included in the centroid calculation (for 0\,\%, the peak position is not corrected at all). If the height becomes too large, the peak tip becomes less symmetric (in case of closely neighboring peaks) and the centroid calculation does not yield the correct peak position. For a peak tip height of 6\,\%, the standard deviation becomes smallest ($2.4^{\circ}$) with an absolute mean of $1.5^{\circ}$. 

When evaluating the SLI line profiles, the determined peak positions were therefore corrected with correction method (iv), using a peak tip height of 6\,\% of the total signal amplitude.


\section{Comparison of SLI and scatterometry profiles}
\label{sec:samples}

\begin{figure}[htbp]
	\centering
	\includegraphics[width=0.8\textwidth]{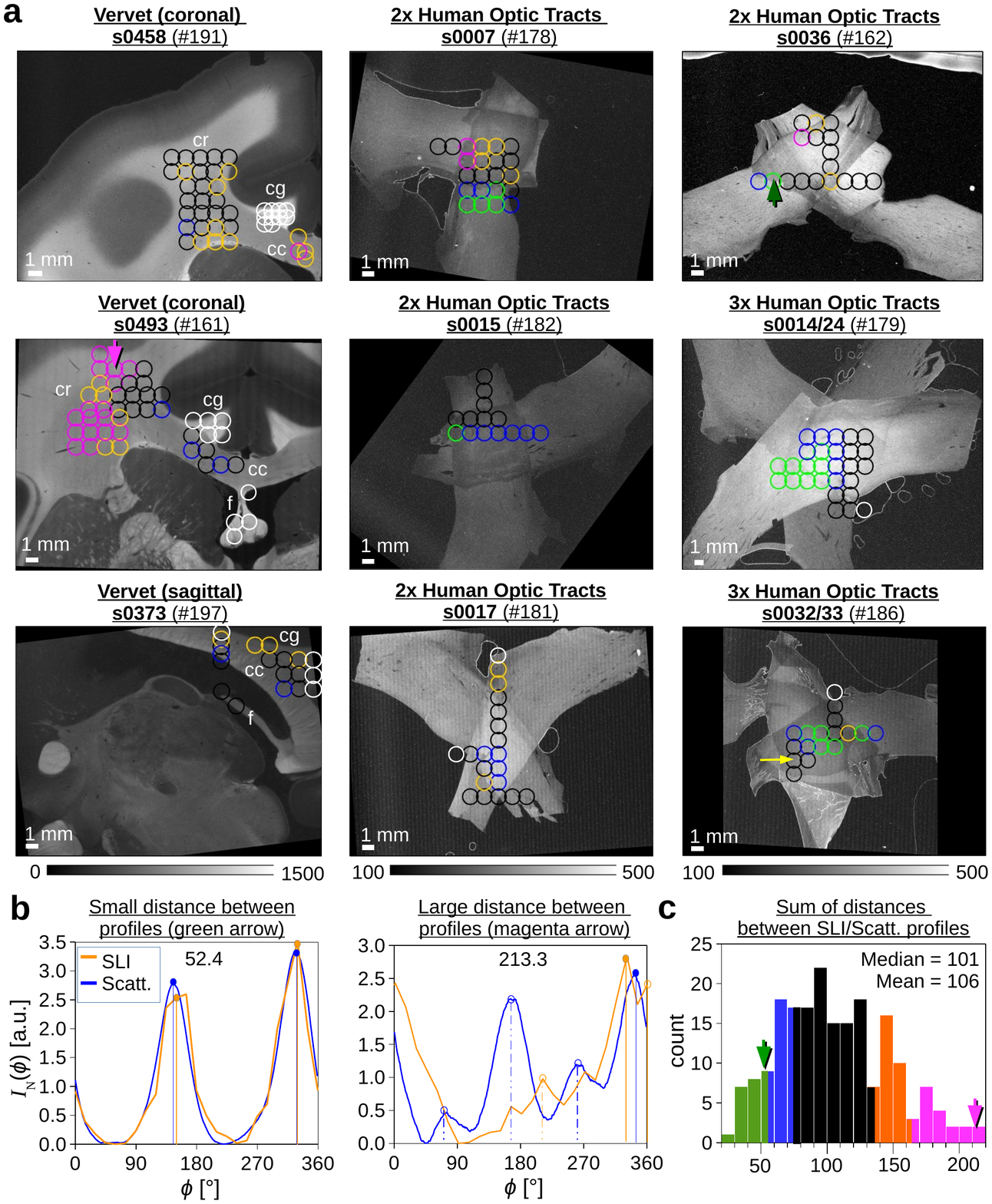}
	\caption{\textbf{Comparison of SLI and scatterometry profiles for different brain tissue samples} (coronal/sagittal vervet brain sections, two/three crossing sections of human optic tracts). \textbf{(a)} Average transmitted light intensity of the SLI measurements, labeled by the sample, section number, and measurement ID\#. \Cref{tab:table} contains more details about the measurement, like the rotation angle by which the SLI images were rotated with respect to the scatterometry measurement. The circles indicate the positions of the tissue spots (with 1.12\,mm diameter) that were measured by scatterometry. In the vervet brain sections, anatomical regions are labeled (cr = corona radiata; cg = cingulum; cc = corpus callosum; f = fornix). \textbf{(b)} Normalized SLI profiles (orange) and scatterometry profiles (blue) obtained from the tissue spots indicated by the green and magenta arrow in (a). The numbers at the top show the sum of distances between SLI and scatterometry profiles. The vertical orange/blue lines indicate the determined peak positions, the dashed-dotted lines the peaks that were not used for comparison. \textbf{(c)} Sum of distances between SLI and scatterometry profiles evaluated for 200 tissue spots (non-white circles in (a)): best 10\,\% with smallest distance (green), best 10--25\,\% (blue), worst 25-10\,\% (orange), worst 10\,\% (magenta).}
	\label{fig:samples}
\end{figure}

To enable a direct comparison of SLI and scatterometry profiles, both profiles need to be generated from exactly the same tissue spot.
In the scatterometry measurement, marker dots were placed on the brain section to determine the initial position of the laser beam during the measurement, micrometer screws were used to move the sample in the x/y-direction in steps of 0.5\,mm or 1\,mm, and the positions of the measured tissue spots were determined by imaging the marker dots on the sample with a digital microscope (see \cite{menzel2020-BOEx} for more details).
To evaluate the same tissue regions with SLI, the image of the average transmitted light intensity $\langle I(\phi) \rangle$ in the SLI measurement was aligned with the corresponding image of the digital microscope, and the determined positions of the tissue spots were transferred to the SLI image. First, the alignment was done manually using anatomical structures as reference, and the SLI profiles were computed by averaging the transmitted light intensity in each 1.12\,mm tissue spot and for each angle $\phi$ in the SLI measurement.
To improve the alignment of the images, the determined peak positions of the SLI profiles were compared to the peak positions of the corresponding scatterometry profiles in order to determine systematic errors. To ensure that only correct peak pairs are compared to each other, the prominent peaks in the scatterometry and SLI profiles were only compared to each other if both line profiles have the same number of prominent peaks (prominence $\geq 8$\,\%). When the difference between the peak positions showed a significant systematic shift for all line profiles in one brain section, \ie\ the mean is not compatible with zero ($|\rm mean| > 3 \sigma / \sqrt{N}$, with standard deviation $\sigma$ and number of peak differences $N$), the SLI images of the brain section were rotated by this value. The positions of the scatterometry spots in the SLI images were corrected accordingly.
\Cref{fig:samples} shows all evaluated spots in the brain sections after the correction (\cref{tab:table} lists the rotation angles $\rho$ by which the SLI images were rotated with respect to the scatterometry measurements). All tissue spots that were used for comparison are marked by a circle (diameter: 1.12\,mm). The non-white circles indicate the tissue spots for which the differences between SLI and scatterometry profiles were computed (see \cref{sec:SLI-scatterometry-comparison}).

As the peaks of the scatterometry profiles were shown to be reliable \cite{menzel2020-BOEx}, also smaller peaks (with prominence $\geq 3\,\%$) were considered for the comparison. For the SLI profiles, only peaks with prominence $\geq 8\,\%$ were considered. To increase statistics, all SLI and scatterometry profiles were compared to each other, even if the number of peaks differs. To ensure that only peaks belonging to the same structures were compared to each other, the peaks were matched by maximizing the sum of inverse squared differences between SLI and scatterometry profiles. That is, the matching of peaks was optimized by putting more weight on peak pairs with a small difference in peak position. Left-over peaks (without partner) were not considered (cf.\ dashed vertical lines in \cref{fig:samples}(b)).
In total, 523 peaks could be compared to each other.
Apart from the difference between peak number and position, we also determined the sum of absolute distances between the normalized SLI and scatterometry profiles $\sum_{\phi} \vert I_{\rm N}(\phi)_{\rm SLI} - I_{\rm N}(\phi)_{\rm Scatt} \vert$ (in steps of $\Delta\phi = 1^{\circ}$, SLI linearly interpolated) to study the overlap between the two curves, see \cref{fig:samples}. The non-white circles indicate how large the differences are (green: best 10\,\% with smallest distance; blue: best 10-25\,\%; orange: worst 25-10\,\%, magenta: worst 10\,\%).

\section{SLI profiles of parallel and crossing nerve fibers}
\label{sec:fiber-crossings}

\begin{figure}[b]
	\centering
	\includegraphics[width=0.9\textwidth]{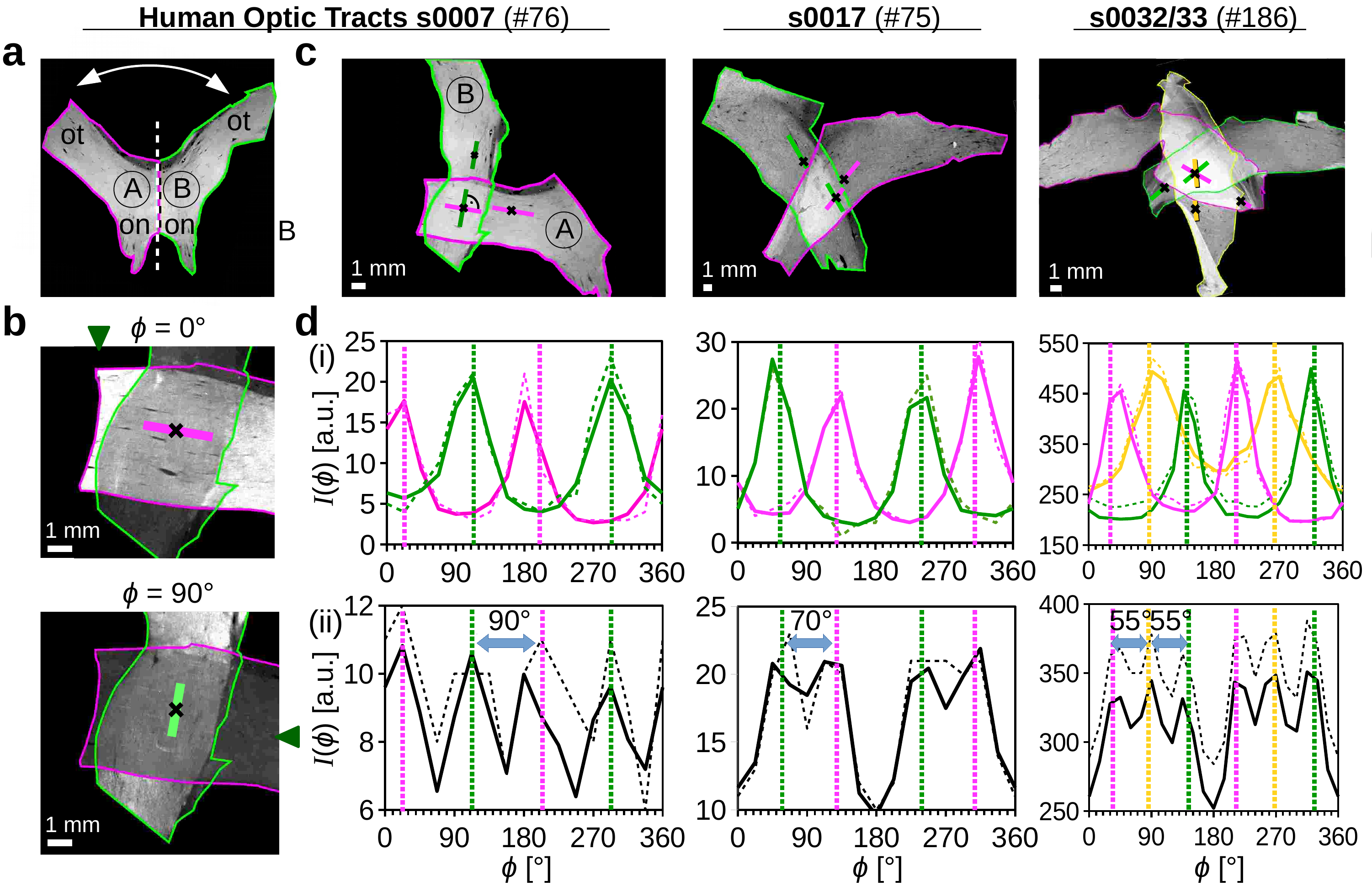}
	\caption{\textbf{SLI profiles for two and three crossing sections of human optic tracts evaluated in regions with parallel and crossing nerve fibers.} \textbf{(a)} Sections of the human optic chiasm were divided into two parts (marked by green and magenta contour lines, A and B) and the sections of the optic tracts (ot) were placed on top of each other under different crossing angles (see (c)). (on = optic nerve) \textbf{(b)} Scattered light intensity during an SLI measurement for a direction of illumination of $\phi = 0^{\circ}$ and $90^{\circ}$. The green arrowheads at the image borders indicate the illumination angle. The straight magenta and green lines indicate the orientations of the fiber tracts which become visible when being illuminated from the broadside. \textbf{(c)} Average transmitted light intensity of an SLI measurement for two and three crossing sections of optic tracts. For better reference, the different sections are delineated by colored contour lines. The black crosses indicate the positions of the evaluated regions, the straight colored lines the predominant orientations of the nerve fibers in the respective regions. \textbf{(d)} Raw SLI profiles $I(\phi)$ obtained from the regions shown in (c), averaged for regions of $10 \times 10$ pixels (solid curves) and for a representative single pixel in their centers (dashed curves). The graphs in (i) belong to regions with non-crossing nerve fibers in the single layer tissue of the optic tract, the graphs in (ii) to regions with crossing nerve fibers in the dual or triple layer overlap. The vertical dashed lines in green/magenta/yellow indicate the approximate positions of the peaks in the SLI profiles in (i) and (ii). The samples with two crossing sections of optic tracts were measured 5 months and the sample with three crossing sections 3 months after tissue embedding (see \cref{tab:table}\#76,75,186).}
	\label{fig:chiasm-parallel-cross-lineprofiles}
\end{figure}

\Cref{fig:chiasm-parallel-cross-lineprofiles}(d) shows the SLI profiles obtained from regions with parallel (i) and crossing (ii) in-plane nerve fibers in two and three crossing sections of human optic tracts. 
The solid curves were obtained from regions with $10 \times 10$ pixels, the dashed curves from the corresponding center of $1 \times 1$ pixels (indicated by small black crosses in (c)). The tiny mismatch between the solid and dashed curves demonstrates the conformity of scattering in  pixels from such areas allowing for the selection of small representative regions.
In regions with parallel in-plane nerve fibers (i), the SLI profiles show two distinct peaks that lie approximately $180^{\circ}$ apart;
the arithmetic mean value of the peak positions corresponds to the directions of the fibers in the section of the optic tract (indicated by green/magenta/yellow lines in (c)). The images in (b) show that one tract lights up when being illuminated from the broadside, thus revealing its underlying structure. 
In regions with crossing  in-plane nerve fibers (iii), the SLI profiles show several distinct peaks, where each $180^{\circ}$-peak pair corresponds to one fiber tract direction. 
The crossing angles of all fiber tracts can be correctly determined in the crossing region, 
not only for two crossing sections of optic tracts with $90^{\circ}$ and $70^{\circ}$ crossing angles but also for three crossing sections with $55^{\circ}$ crossing angle. 
A comparison between the graphs in (i) and (ii) shows that the signal in the crossing region is a superposition of the signals in the corresponding single tissue layer: The peaks in (i) and (ii) occur at similar positions, see vertical dashed lines.


\section{Time-dependent changes in scattering}
\label{sec:long-term}

\begin{figure}[htbp]
	\includegraphics[width=0.8\textwidth]{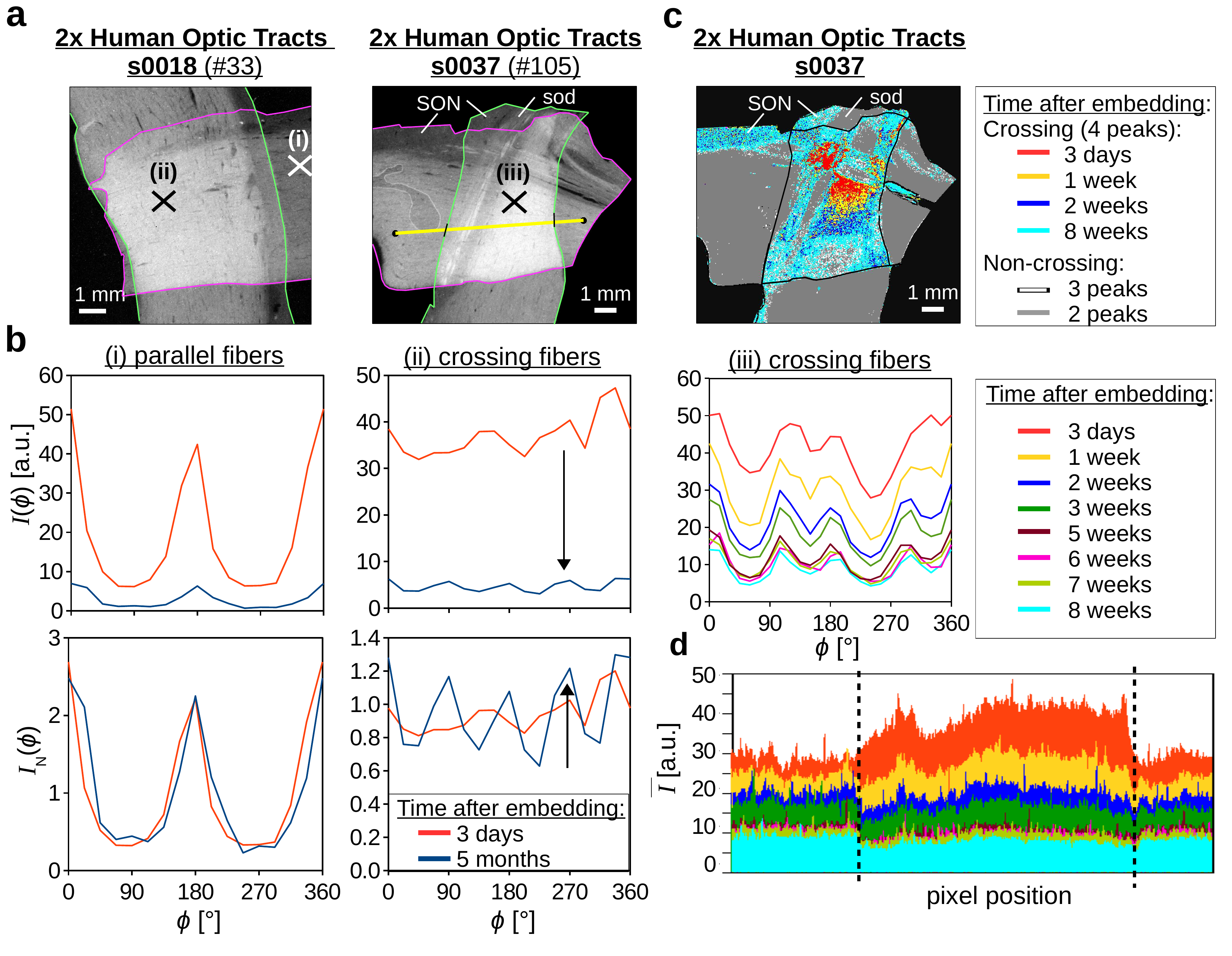}
	\caption{\textbf{Time-dependent changes in scattering for two crossing sections of human optic tracts} (sections 18 and 37, \cref{tab:table}\#33,105). \textbf{(a)} Average transmitted light intensity of the SLI measurement 3 days after tissue embedding. Three regions with $5 \times 5$ pixels were selected for evaluation (marked by crosses). The different sections of the optic tracts were delineated by magenta/green contour lines for better reference. (SON = supraoptic nucleus, sod = supraoptic decussation) \textbf{(b)} Non-normalized SLI profiles $I(\phi)$ and normalized SLI profiles $I_{\rm N}(\phi)$ for the three selected regions, evaluated at different times after tissue embedding (3 days up to 5 months). \textbf{(c)} Brain areas for which the SLI profiles show four prominent peaks (interpreted as two crossing fiber bundles) are displayed in different colors, depending on the time elapsed after tissue embedding. Areas for which the SLI profiles show two prominent peaks (interpreted as in-plane parallel fibers) are shown in gray. \textbf{(d)} Intensity values of the average transmitted light intensity of the SLI measurement evaluated along the yellow line in (a) for measurements at eight different times after tissue embedding (3 days until 8 weeks). The vertical dashed lines indicate the borders of the crossing region.}
	\label{fig:chiasm-longterm}
\end{figure}

To increase birefringence and scattering contrast, all brain tissue samples were embedded in a glycerin solution before being cover-slipped (see \cref{sec:preparation}).
To study the effect of embedding time on the scattering properties of brain tissue, the same brain tissue samples were measured at different times after embedding (from 3 days up to 8 week). \Cref{fig:chiasm-longterm}(b) shows the (non-)normalized SLI profiles for two crossing sections of optic tracts, evaluated for a region with parallel in-plane fibers (i) and two regions with crossing in-plane fibers ((ii),(iii)) obtained from SLI measurements at different times after embedding. 

With increasing time elapsed after tissue embedding, the scattering decreases notably, especially in the first few weeks (see (iii)). Within 8 weeks, the average light intensity of the SLI signal decreases by more than 70\,\%, after 5 months it is reduced by more than 85\,\%. \Cref{fig:chiasm-longterm}(d) shows the average intensity  evaluated along a geometric line profile (yellow line in (a)). The vertical dashed lines indicate the borders of the crossing region. While the crossing region (double tissue layers) shows stronger scattering than the region with parallel fibers (single tissue layer) right after the tissue embedding, it is the other way around after 2 weeks. 
Especially interesting is the comparison of the normalized SLI profiles: While the profiles in regions with parallel in-plane fibers remain almost the same (see lower image in (i)), the peaks in regions with crossing  in-plane fibers become more prominent with increasing time after tissue embedding (see lower image in (ii)), \ie\ the amplitude of the intensity profile does not scale down proportionally to the lowering of its average, but less. \Cref{fig:chiasm-longterm}(c) shows the brain areas for which four peaks are detected (interpreted as two in-plane crossing fiber bundles) in different colors, depending on the time elapsed after tissue embedding. Three days after tissue embedding, only a small part within the crossing region is correctly classified as two crossing fiber bundles (in red). This region grows as time goes by (yellow and blue). After 8 weeks, four peaks are detected in the majority of the pixels in the crossing region (cyan), but also in the gray matter (SON). 
One possible explanation why the fiber directions in regions with in-plane crossing fibers are more correctly determined for brain sections with longer embedding time and reduced scattering might be that strongly scattering fiber bundles influence each other (higher-order scattering), reducing the prominence/contrast of the resulting peaks in the SLI profile. 

As the prominence of the peaks in the SLI profiles is a crucial criterion for a correct peak detection, a longer lapse of time between embedding and measurement leads to a more reliable fiber crossing classification even at a lower level of scattering (see \cref{fig:chiasm-longterm}(b)(ii)). However, it should be noted that a long embedding time might also lead to tissue deformation and increases the transmittance so that a combined measurement of SLI and 3D-PLI is no longer possible as the transmittance is used in 3D-PLI to correct the determined fiber inclinations.


\section{Supplementary table and figures}
\label{sec:suppl-images}

{ 
	\renewcommand{\arraystretch}{1.4}
	\renewcommand{\tabcolsep}{1mm}
	\begin{table}[h]
		\footnotesize
		\begin{center}
			\begin{tabular}{ c  c  c  c  c  c  c  c  c  c  c  c  c }
				\hline
				\textbf{sample}		& \textbf{section}	& \textbf{$T$} 	& \textbf{\#} 	& \textbf{SLI} 	& \textbf{scatt.}  & $\rho$ 	& \textbf{mask}	& $f$ 	&$h$ 	& $l$ 	& $t$ & px 	\\ \addlinespace[-0.1cm]
				& \textbf{no.}		& [\um]			&   	& [days]		& [days]			& [$^{\circ}$] &					& [mm]	& [cm]	& [cm]	& [s] & [\um]	\\ \hline\hline
				\multirow{2}{*}{\shortstack[|c|]{\textbf{Wistar Rat} \\[0.8mm] (coronal)}} 					
				& 	\multirow{2}{*}{\textbf{157}}		& 	\multirow{2}{*}{60} 			& 	\multirow{2}{*}{\textbf{78}}  	& 	\multirow{2}{*}{1}				& 	\multirow{2}{*}{--}		& 	\multirow{2}{*}{--}			& 		\multirow{2}{*}{rect}	& 		\multirow{2}{*}{90}	& 		\multirow{2}{*}{10.9}	&  	\multirow{2}{*}{27.8} & 	\multirow{2}{*}{0.5} & 	\multirow{2}{*}{6.5}	\\ 
				& & & & & & & & & & & & \\ \hline
				\multirow{2}{*}{\shortstack[|c|]{\textbf{Vervet} \\[0.8mm] (sagittal)}}				
				& \textbf{373}		& 60 			& \textbf{197}  & 105*			& 118				& 	6.9 & rect	& 	90	& 	10.9	&  27.8 & 1.0	& 6.5	\\ \cline{2-13}
				& \textbf{374}		& 60 			& \textbf{139}  & 5*			& --				& 	--  & rect	& 	50	& 	10.9	&  28.3	& 0.5	& 15.0 \\ \hline
				\multirow{3}{*}{\shortstack[|c|]{\textbf{Vervet} \\[0.8mm] (coronal)}} 				
				& \textbf{458}		& 60 			& \textbf{191}  & 13			& 22				&	0.0 & rect	&	90	& 	10.9	&  27.8 & 1.0	& 6.5	\\ \cline{2-13}
				& \textbf{493} 		& 60 			& \textbf{161}  & 9				& 43 				&  	1.5 & rect	& 	90	& 	10.9	&  27.8 & 2.0	& 6.5	\\ \cline{2-13}
				& \textbf{512} 		& 60 			& \textbf{90}  	& 1				& -- 				&  	--  & circ	& 	50	& 	10.9	&  19.3 & 0.5	& 13.7	\\ \hline
				\multirow{13}{*}{\shortstack[|c|]{\textbf{Human} \\[0.8mm] (optic tracts)}} 			
				& \multirow{2}{*}{\textbf{7}}	& \multirow{2}{*}{60}	& \textbf{76}	& 153			& --				& 	--  & circ	& 	90	& 	10.9	&  33.8 & 0.5 & 8.3		\\ \cline{4-13}
				& 					& 				& \textbf{178}	& 275			& 293				& 	-9.8 & rect	& 	90	& 	10.9	&  27.8 & 1.0	& 6.5	\\ \cline{2-13}
				& \textbf{15}		& 30			& \textbf{182}	& 275			& 293				& 	53.4 & rect	& 	90	& 	10.9	&  27.8 & 1.0	& 6.5	\\ \cline{2-13}
				& \multirow{2}{*}{\textbf{17}}	& \multirow{2}{*}{30}	& \textbf{75}	& 153			& --				& 	--  & circ	& 	90	& 	10.9	&  34.5 & 0.5 &	8.8	\\ \cline{4-13}
				& 					& 				& \textbf{181}	& 275			& 294				& 	4.0 & rect	& 	90	& 	10.9	&  27.8 & 1.0 & 6.5		\\ \cline{2-13}
				& \multirow{2}{*}{\textbf{18}}	& \multirow{2}{*}{30} & \textbf{33}	& 3		& --		& 	--  & circ	& 	90	& 	11.0	&  33.7 & 0.5 &	8.6	\\ \cline{4-13}
				&					&				& \textbf{74}	& 153			& --				&   --  & circ	&	90	&	10.9	&  34.8 & 0.5 & 8.8		\\ \cline{2-13}
				& \textbf{14/24}	& 30			& \textbf{179}	& 275			& 294				& 	0.0 & rect	& 	90	& 	10.9	&  27.8 & 1.0	& 6.5	\\ \cline{2-13}
				& \textbf{32/33}	& 30			& \textbf{186}	& 112			& 126				& 	-3.4 & rect	&	90	& 	10.9	&  27.8 & 1.0	& 6.5	\\ \cline{2-13}
				& \multirow{3}{*}{\textbf{36}}	& \multirow{3}{*}{30} & \textbf{106} & 0			& --				& 	--  & rect	&	90	& 	10.9	&  33.8 & 0.5 &	8.3	\\ \cline{4-13}
				& 				& 					&\textbf{158} 	& 2*			& --				& 	--  & rect	&	90	& 	10.9	&  27.8 & 2.0	& 6.5	\\ \cline{4-13}
				&								&					& \textbf{162}	& 9*			& 43				& 	0.5 & rect	& 	90	& 	10.9	&  27.8 & 2.0	& 6.5	\\ \cline{2-13}
				& \textbf{37}	& 30				& \textbf{105}	& 3				& --				& 	--  & rect	&	90	&	10.9	&  33.8	& 0.5 & 8.3 \\ \hline
			\end{tabular}
		\end{center}
		\caption{\textbf{List of sample properties and measurement parameters for all investigated brain tissue samples:} sample, section number, section thickness ($T$), measurement ID (\#), dates of SLI and scatterometry measurements (in days after tissue embedding), rotation angle ($\rho$) by which the SLI images were rotated with respect to the scatterometry measurement (see \cref{fig:samples}); settings used for the SLI measurements (see \cref{fig:scattering-setup}): mask (circular/rectangular), focal length of camera objective ($f$), distance between light source and sample ($h$), distance between sample and camera objective ($l$), exposure time ($t$), and pixel size in object space (px). The asterisk indicates that the given number of days are the days after the revitalization of the tissue (not after the initial embedding).}
		\label{tab:table}
	\end{table}
}

\begin{figure}[htb]
	\centering
	\includegraphics[width=0.7\textwidth]{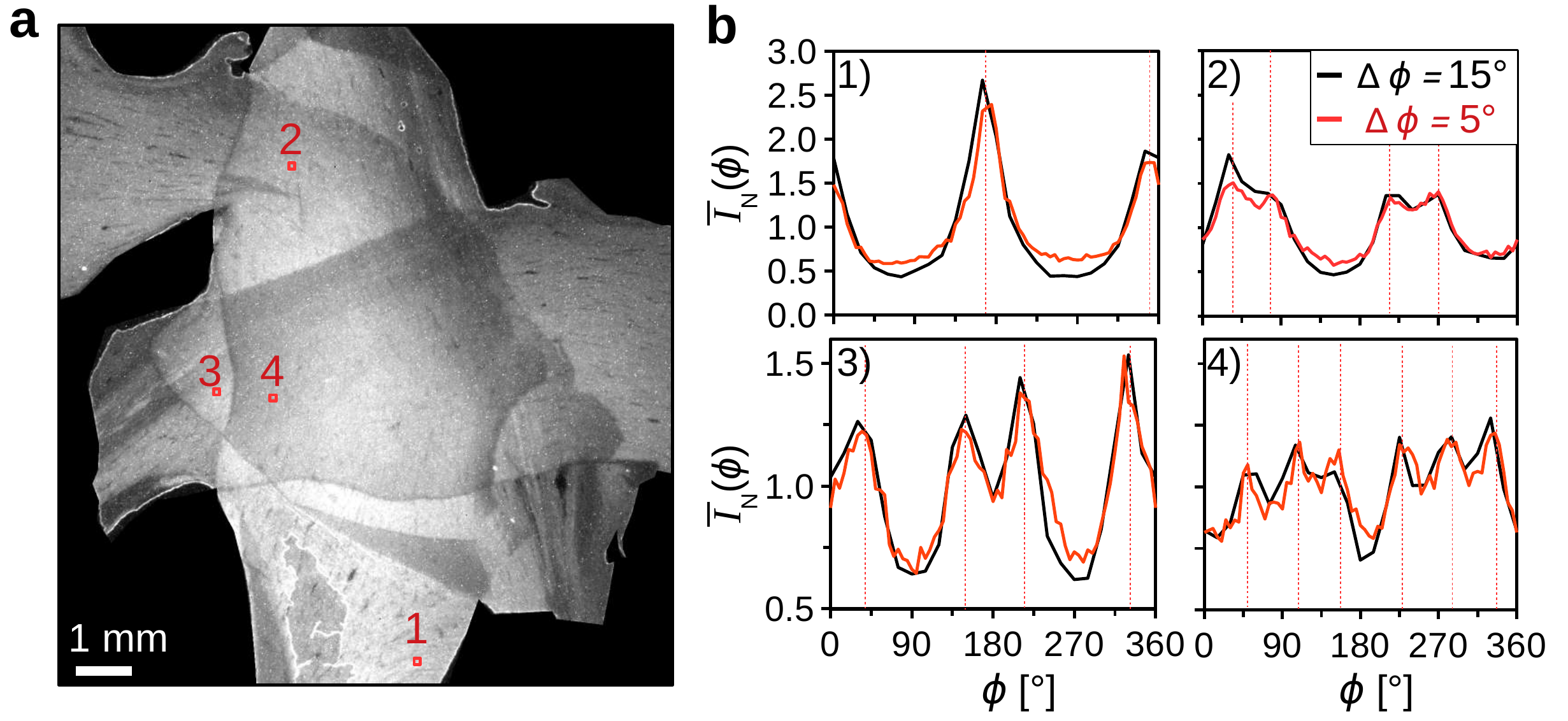}
	\caption{\textbf{SLI measurement with $15^{\circ}$ and $5^{\circ}$-steps for three crossing sections of human optic tracts} (see \cref{tab:table}\#143). \textbf{(a)} Average of the transmitted light intensity in the SLI measurement (5 months after embedding) localizing four selected regions of interest (10 $\times$ 10 pixels). \textbf{(b)} SLI profiles of the selected regions obtained from a measurement with $15^{\circ}$-steps (black curves) and $5^{\circ}$-steps (red curves). The red vertical lines indicate the peak positions of the red curves. The measurements were performed with the masks with rectangular holes described in \cref{sec:SLI}. The peak positions obtained from the $15^{\circ}$ measurement are very similar to those obtained from the $5^{\circ}$ measurement.}
	\label{fig:chiasm-3xcross_15-vs-5deg}
\end{figure}

\begin{figure}[htb]
	\centering
	\includegraphics[width=0.6\textwidth]{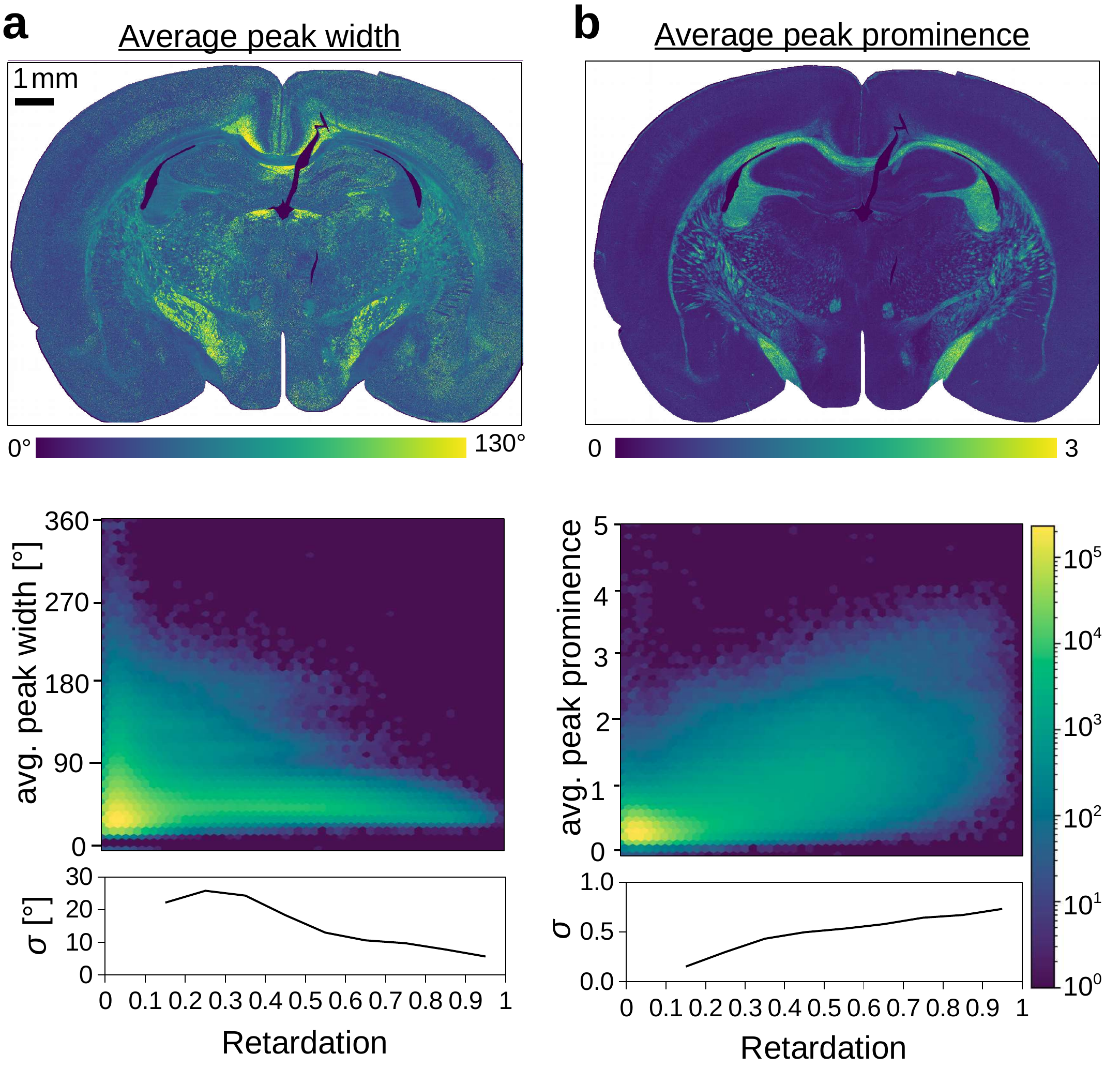}
	\caption{\textbf{Average peak width and prominence of the coronal rat brain section} (\cref{tab:table}\#78). For peaks with a prominence $\geq 8\,\%$ of the total signal amplitude, the peak width and prominence were computed from the normalized SLI profiles for each image pixel and averaged for each SLI profile. The 2D histograms show the average peak width and prominence plotted against the retardation for each image pixel. The graphs below show the standard deviation $\sigma$ plotted against the retardation.}
	\label{fig:rat_supplement}
\end{figure}

\begin{figure}[htb]
	\centering
	\includegraphics[width=0.9\textwidth]{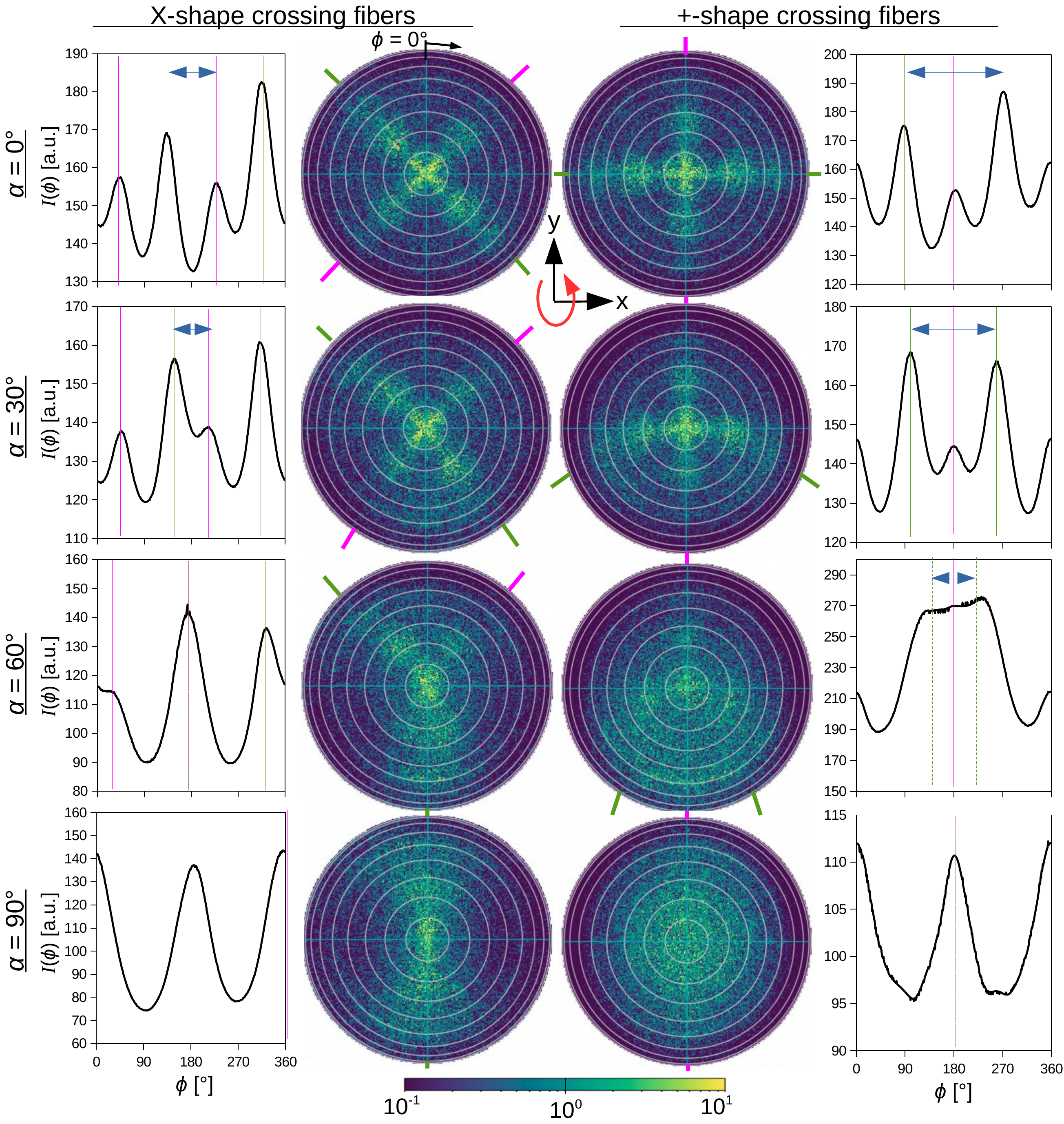}
	\caption{\textbf{Simulated scattering patterns for $90^{\circ}$-crossing, inclined fiber bundles.} X-shape crossing fiber bundles (oriented diagonally to the x/y-axis) and $+$-shape crossing fiber bundles (oriented along x and y) were rotated around the x-axis (as indicated by the red arrow) to obtain crossing fiber bundles with different inclination angles $\alpha$. The middle images show the corresponding scattering patterns, the graphs at the left/right show the line profiles obtained after performing a Gaussian blur with $8^{\circ}$ radius (cf.\ \cref{sec:simulation}) and computing the radial sum of the scattering patterns. Scattering peaks are indicated by green/magenta lines. The simulations were performed as described in \cite{menzel2020}, for $90^{\circ}$-crossing interwoven fibers in a volume of $50 \times 50 \times 50$\,\textmu m$^3$. For the X-shaped crossing fibers (left), the scattering peaks in the lower half of the scattering pattern merge much faster than the scattering peaks in the upper half of the scattering pattern with increasing fiber inclination. The scattering pattern of the +-shaped crossing fibers (right) is a superposition of the scattering pattern of an in-plane parallel fiber bundle (x-bundle) and an inclined fiber bundle (y-bundle): While the peak positions of the x-bundle (magenta peaks) remain unchanged, the peaks of the y-bundle (green peaks) merge with increasing fiber inclination as expected for inclined fibers. The study shows that line profiles of strongly inclined crossing fibers look different from line profiles of in-plane crossing or parallel inclined fibers and might be used for distinction.}
	\label{fig:cross_incl}
\end{figure}

\cleardoublepage
\bibliography{BIBLIOGRAPHY}


\end{document}